\input harvmac.tex
%\draftmode
%%%%%%%%%%%  Math-style letters   %%%%%%%%%
\font\cmss=cmss10 \font\cmsss=cmss10 at 7pt

\def\Z{\relax\ifmmode\mathchoice
{\hbox{\cmss Z\kern-.4em Z}}{\hbox{\cmss Z\kern-.4em Z}}
{\lower.9pt\hbox{\cmsss Z\kern-.4em Z}}
{\lower1.2pt\hbox{\cmsss Z\kern-.4em Z}}\else{\cmss Z\kern-.4em
Z}\fi}
\def\IZ{Z\!\!\!Z}
\def\C{{\bf C}}
\def\H{{\bf H}}

\def\R{{\bf R}}
\def\tilde{\widetilde}

\def\bar{\overline}

%%%%%%%%%%%%% Calligraphic letters  %%%%%%

\def\CM {{\cal M}}
\def\CN {{\cal N}}
\def\CO {{\cal O}}

\lref\brmi{R.~Britto-Pacumio, J.~Michelson, A.~Strominger and
 A.~Volovich, ``Lectures on superconformal quantum mechanics and
 multi-black hole  moduli spaces'', hep-th/9911066.}
\lref\fegi{S. Ferrara, G. Gibbons and R. Kallosh, ``Black holes and
 critical points in moduli space'', hep-th/9702103.}
\lref\witten{E. Witten, ``Supersymmetry and Morse theory'',
 J.Diff.Geom. {\bf 17}(1982) 661.}
\lref\abto{E.~R.~Abraham and P.~K.~Townsend,
 ``Q kinks'',
 Phys.\ Lett.\  {\bf B291}, 85 (1992).}
\lref\baho{C. Bachas, J. Hoppe and B. Pioline, ``Nahm equations,
 $N=1^*$ domain walls, and D-strings in $AdS_5 \times S_5$''
 hep-th/0007067.}
\lref\luov{A.~Lukas, B.~A.~Ovrut, K.~S.~Stelle and D.~Waldram,
 ``Heterotic M-theory in five dimensions,'' Nucl.\ Phys.\  {\bf B552},
 246 (1999), hep-th/9806051.}
\lref\begu{K. Behrndt and S. Gukov, ``Domain walls and
 superpotentials from {M} theory on {C}alabi-{Y}au three-folds'',
 Nucl.\ Phys.\  {\bf B580}, 225 (2000), hep-th/0001082.}
\lref\gusi{M.~Gunaydin, G.~Sierra and P.~K.~Townsend,
 ``Gauging The D = 5 Maxwell-Einstein Supergravity Theories:
 More On Jordan Algebras'', Nucl.\ Phys.\  {\bf B253}, 573 (1985).}
\lref\den{F.\ Denef, ``Supergravity flows and D-brane stability'',
 hep-th/0005049}
\lref\feka{S.\ Ferrara, R.\ Kallosh and A.\ Strominger,
 ``N=2 extremal black holes,'' Phys.\ Rev.\  {\bf D52} (1995) 5412,
 hep-th/9508072}
\lref\fekaa{S.\ Ferrara and R.\ Kallosh,
 ``Supersymmetry and attractors,'' Phys.\ Rev.\  {\bf D54} (1996) 1514,
 hep-th/9602136}
\lref\cvgr{M.\ Cveti{\v c} and S.\ Griffies, ``Domain walls in N=1
 supergravity'', hep-th/9209117;
 M.~Cvetic and H.~H.~Soleng, ``Supergravity domain walls,''
 Phys.\ Rept.\  {\bf 282} (1997) 159, [hep-th/9604090].}
\lref\frgu{D.Z.\ Freedman, S.S.\ Gubser, K.\ Pilch and N.P.\ Warner,
 ``Renormalization group flow from holography, supersymmetry
  and a c-theorem'', hep-th/9904017.}
\lref\be{K.\ Behrndt, ``Domain walls of D = 5 supergravity and fixed
 points of N = 1 super  Yang-Mills,'' Nucl.\ Phys.\  {\bf B573}
 (2000) 127, hep-th/9907070.}
\lref\hesk{M.\ Henningson and K.\ Skenderis, ``Holography and
 the Weyl anomaly''JHEP {\bf 9807}, 023 (1998), hep-th/9812032.}
\lref\algo{E.\ Alvarez and C.\ Gomez, ``Geometric holography,
 the renormalization group and the c-theorem,''
 Nucl.\ Phys.\  {\bf B541} (1999) 441, hep-th/9807226.}
\lref\gipe{L.~Girardello, M.~Petrini, M.~Porrati and A.~Zaffaroni,
 ``Novel local CFT and exact results on perturbations of N = 4 super
 Yang-Mills from AdS dynamics,'' JHEP {\bf 9812} (1998) 022
 hep-th/9810126.}
\lref\gusii{M.~Gunaydin, G.~Sierra and P.~K.~Townsend,
 ``The geometry of N=2 Maxwell-Einstein supergravity and Jordan
 algebras'', Nucl.\ Phys.\  {\bf B242}, 244 (1984).}
\lref\greene{B.~R.~Greene, K.~Schalm and G.~Shiu, ``Dynamical topology
 change in M theory,'' hep-th/0010207.}
\lref\groupsref{M.~ Gunaydin, M.~ Zagermann, ``The Gauging of
Five-dimensional, N=2 Maxwell-Einstein Supergravity Theories coupled to Tensor
Multiplets,'' Nucl.Phys. {\bf B572} (2000) 131.}
\lref\ganor{Y.-K. E.~ Cheung, O.J.~Ganor, M.~ Krogh,
``On the Twisted (2,0) and Little String Theories,''
Nucl. Phys. {\bf B536} (1998) 175, hep-th/9805045.}
\lref\borel{A.~Borel, ``Sur la cohomologie des espaces fibr\'es
principaux
et des espaces homog\`enes de groupes de Lie compacts,''
Ann. of Math. (2) {\bf 57} (1953) 115;
A.~Borel, F.~Hirzebruch, ``Characteristic Classes
and Homogeneous Spaces -- I, II, III'', Amer. J. Math. {\bf 80} (1958)
458;
Amer. J. Math. {\bf 81} (1959) 315; Amer. J. Math. {\bf 82} (1960) 491.}

\lref\onishchic{A.~L.~Onishchic, ``Topology of Transitive Transformation
Groups'', 1994.}
\lref\gorenstein{D.~Gorenstein, ``Finite Simple Groups, An Introduction
to Their Classification'', Plenum Press, New York and London, 1982.}

\lref\town{P.K. Townsend, ``Positive energy and the scalar potential
 in higher dimensional (super) gravity'', Phys.\ Lett.\ {\bf
 B148}, 55 (1984).}
\lref\anbe{L.~Andrianopoli, M.~Bertolini, A.~Ceresole, R.~D'Auria,
 S.~Ferrara, P.~Fre and T.~Magri, ``N = 2 supergravity and N = 2
 super Yang-Mills theory on general scalar  manifolds:
 Symplectic covariance, gaugings and the momentum map,''
 J.\ Geom.\ Phys.\  {\bf 23}, 111 (1997)
 hep-th/9605032.}
\lref\ceda{A.~Ceresole and G.~Dall'Agata, ``General matter coupled N =2,
 D = 5 gauged supergravity,'' hep-th/0004111.}
\lref\behe{K.~Behrndt, C.~Herrmann, J.~Louis and S.~Thomas,
 ``Domain walls in five dimensional supergravity with
 non-trivial  hypermultiplets,'' hep-th/0008112.}
\lref\kali{R.~Kallosh and A.~Linde, ``Supersymmetry and the brane
world,''
 JHEP {\bf 0002}, 005 (2000),hep-th/0001071.}
\lref\cace{A.~C.~Cadavid, A.~Ceresole, R.~D'Auria and S.~Ferrara,
 ``Eleven-dimensional supergravity compactified on Calabi-Yau
 threefolds,'' Phys.\ Lett.\  {\bf B357}, 76 (1995), hep-th/9506144.}
\lref\wipr{B.~de Wit and A.~Van Proeyen, ``Isometries of special
manifolds,''
 hep-th/9505097}
\lref\guza{M.~Gunaydin and M.~Zagermann, ``The gauging of
five-dimensional,
 N = 2 Maxwell-Einstein supergravity  theories coupled to tensor
 multiplets,'' Nucl.\ Phys.\  {\bf B572} (2000) 131, hep-th/9912027;
 ``The vacua of 5d, N = 2 gauged
 Yang-Mills/Einstein/tensor supergravity:  Abelian case,''
 Phys.\ Rev.\  {\bf D62} (2000) 044028, hep-th/0002228.}
\lref\wila{B.~de Wit, P.~G.~Lauwers and A.~Van Proeyen,
 ``Lagrangians Of N=2 Supergravity - Matter Systems,''
 Nucl.\ Phys.\  {\bf B255}, 569 (1985).}
\lref\kama{A.~Chou, R.~Kallosh, J.~Rahmfeld, S.~Rey,
 M.~Shmakova and W.~K.~Wong, ``Critical points and phase transitions
 in 5d compactifications of  M-theory,'' Nucl.\ Phys.\  {\bf B508},
 147 (1997), hep-th/9704142.}
\lref\becvv{K.~Behrndt and M.~Cvetic, ``Anti-de Sitter vacua of gauged
 supergravities with 8 supercharges,'' Phys.\ Rev.\  {\bf D61},
 101901 (2000), hep-th/0001159.}
\lref\tava{T.~R.~Taylor and C.~Vafa, ``RR flux on Calabi-Yau and partial
 supersymmetry breaking,'' Phys.\ Lett.\  {\bf B474}, 130 (2000),
 hep-th/9912152.}
\lref\mi{J.~Michelson, ``Compactifications of type IIB strings to four
 dimensions with  non-trivial classical potential,''
 Nucl.\ Phys.\  {\bf B495}, 127 (1997), hep-th/9610151.}
\lref\gu{S.~Gukov, ``Solitons, superpotentials and calibrations,''
 Nucl.\ Phys.\  {\bf B574}, 169 (2000), hep-th/9911011.}
\lref\ADS{J.M.~ Maldacena, "The Large N Limit of Superconformal
 Field Theories and Supergravity", Adv. Theor. Math. Phys.
 {\bf 2} (1998) 231; S.S.~ Gubser, I.~R.~ Klebanov, A.~M.~ Polyakov,
 "Gauge Theory Correlators from Non-Critical String Theory",
 Phys. Lett. {\bf B428} (1998) 105; E.~Witten, "Anti De Sitter
 Space And Holography", Adv. Theor. Math. Phys. {\bf 2} (1998) 253.}
\lref\khwa{A.~Khavaev and N.~P.~Warner, ``A class of N = 1
supersymmetric
 RG flows from five-dimensional N = 8  supergravity,'' hep-th/0009159.}
\lref\wit{E.~Witten, ``Phase Transitions In M-Theory And F-Theory,''
 Nucl.\ Phys.\  {\bf B471} (1996) 195, [hep-th/9603150].}
\lref\Aspinwall{P.~Aspinwall, private communication.}
\lref\BST{H.J.~Boonstra, K.~Skenderis, P.K.~Townsend,
``The domain-wall/QFT correspondence", JHEP {\bf 9901} (1999) 003.}
\lref\Moore{G.~Moore, ``Arithmetic and Attractors," hep-th/9807087.}
\lref\kls{R. ~Kallosh, A. ~Linde, M. ~Shmakova, "Supersymmetric
Multiple Basin Attractors", JHEP {\bf 9911} (1999) 010}
\lref\bhk{R. ~Kallosh, "Multivalued Entropy of Supersymmetric Black
Holes,"
JHEP {\bf 0001} (2000) 001}
\lref\FGK{ S. Ferrara, G. W. Gibbons, R. Kallosh, ``Black Holes and Critical
Points in Moduli Space," Nucl.Phys. B500 (1997) 75, hep-th/9702103}
\lref\ddual{I. Antoniadis, S. Ferrara, T.R. Taylor, Nucl.Phys. B460
(1996) 489, hep-th/9511108.}
\lref\PSFL{ J. Polchinski, A. Strominger, Phys.Lett. B388 (1996) 736,
hep-th/9510227.}
\lref\bhsusy{ A. Chamseddine, S. Ferrara, G. Gibbons, R. Kallosh,
Phys.Rev. D55 (1997) 3647, hep-th/9610155.}
\lref\fdbhg{K.~Behrndt, A. H.~Chamseddine and W. A.~Sabra, ``BPS black
holes in
N=2 five dimensional ADS supergravity,'' Phys.\ Lett.\  {\bf
B}422,(1998) 97,
hep-th/9807187.}
\lref\GS{M.~ Gutperle, M.~ Spalinski, ``Supergravity Instantons
for N=2 Hypermultiplets," hep-th/0010192.}
\lref\Sabra{W.~A.~Sabra, ``Black holes in N=2 supergravity theories
and harmonic functions," Nucl.Phys. {\bf B510} (1998) 247,
hep-th/9704147;
``General BPS black holes in five dimensions,''
Mod.\ Phys.\ Lett.\ {\bf A13}, 239 (1998),
hep-th/9708103.}
\lref\CKLT{G.~ Curio, A.~ Klemm, D.~ Luest, S.~ Theisen,
``On the Vacuum Structure of Type II String Compactifications
on Calabi-Yau Spaces with H-Fluxes," hep-th/0012213.}
\lref\progress{work in progress.}
\lref\kbtsing{ E.~ Bergshoeff, R.~ Kallosh, A.~ Van Proeyen,
"Supersymmetry in singular spaces,"  JHEP 0010 (2000) 033,
hep-th/0007044.}
\lref\becve{K.~Behrndt and M.~Cvetic,
	``Supersymmetric domain wall world from D = 5 simple gauged 
	supergravity,''
	Phys.\ Lett.\ {\bf B475}, 253 (2000), hep-th/9909058.}
\lref\becvd{ K.~Behrndt and M.~Cvetic, 
	``Gauging of N = 2 supergravity hypermultiplet and novel 
	renormalization  group flows,''hep-th/0101007.}
\lref\KT{I.R.~ Klebanov, A.A.~ Tseytlin, ``Gravity Duals of Supersymmetric
SU(N) x SU(N+M) Gauge Theories," Nucl.Phys. {\bf B578} (2000) 123.}
\lref\BW{J.~Bagger and E.~Witten, Nucl.Phys. {\bf B222} (1983) 1.}
\lref\NCMorse{S.~Wu, ``On the Instanton Complex of Holomorphic
Morse Theory," math.AG/9806118.}
%M.A.~Shubin, ``Semiclassical asymptotics on covering manifolds
%and Morse inequalities," Geom. Funct. Anal. {\bf 6} (1996) 370.

%%%%%%%%%%%%%%%%%%%%%%%%%%%%%%%%%%%%%%%%%%%%%%%%%%%%%%%%%%%%%%%%%

\Title{\vbox{\baselineskip12pt
\hbox{hep-th/0101119}
\hbox{ITEP-TH-71/00}
\hbox{CALT 68-2306}
\hbox{CITUSC/00-062}
\hbox{HU-EP-00/54}
\hbox{SLAC-PUB-8749}}}
{{\vbox{\centerline{Domain Walls, Black Holes, and}
\medskip
\centerline{Supersymmetric Quantum Mechanics }}}}
\centerline{Klaus Behrndt$^a$\foot{Email: behrndt@physik.hu-berlin.de}\ , \quad
Sergei Gukov$^b$\foot{Email: gukov@theory.caltech.edu} \quad 
and \quad Marina Shmakova$^{c}$\foot{Email: shmakova@slac.stanford.edu}}
\bigskip
%\bigskip

\centerline{$^a$ \it Institut f\"ur Physik, Humboldt Universit\"at, 10115 Berlin, Germany}
\medskip
\centerline{$^b$ \it Department of Physics, Caltech,  Pasadena, CA 91125, USA}
\centerline{\it CIT-USC Center for Theoretical Physics, UCS, Los Angeles, CA 90089, USA}
\medskip
\centerline{$^c$ \it CIPA, 366 Cambridge Avenue Palo Alto, CA 94306, USA}
\centerline{\it Stanford Linear Accelerator Center, Stanford University, Stanford, CA 94309, USA}

\vskip .3in
\centerline{\bf Abstract}

Supersymmetric solutions, such as BPS domain walls or black holes,
in four- and five-dimensional supergravity theories with eight supercharges
can be described by effective quantum mechanics with a potential term.
We show how properties of the latter theory can help us to learn about
the physics of supersymmetric vacua and BPS solutions in these supergravity
theories. The general approach is illustrated in a number of specific examples
where scalar fields of matter multiplets take values in symmetric coset spaces.

\Date{January 2001}

%%%%%%%%%%%%%%%%%%%%%%%%%%%%%%%%%%%%%%%%%%%%%%%%%%%%%%%%%%%%%%%%%%%%%

\newsec{Introduction and Summary}

%%%%%%%%%%%%%%%%%%%%%%%%%%%%%%%%%%%%%%%%%%%%%%%%%%%%%%%%%%%

Supersymmetric vacuum configurations in gauged supergravity
theories in four and five dimensions recently receive a lot
of attention due to their relevance to AdS/CFT correspondence \ADS,
and, more generally, to holography in non-superconformal field
theories and Domain Wall/QFT correspondence \BST.
Of particular interest are BPS solutions which preserve
at least four supersymmetries.

In many of such supergravity solutions, the values of scalars
and other fields depend only on a single spatial coordinate.
For example, in the case of a BPS black hole this is a radial
coordinate, while for a domain wall this is a transverse spatial
coordinate.
In any case, one can integrate out the dynamics in the other
directions which are isometries of the solution and obtain
a 0+1 dimensional theory -- supersymmetric quantum mechanics --
where the distinguished spatial coordinate plays the role of time.
This relation to supersymmetric quantum mechanics was
used in certain supergravity solutions, see {\it e.g.} \KT.
Interpretation of the attractor flow as supersymmetric quantum
mechanics was also suggested in \Moore.
%in a similar, but different, context is very well known in the case
%of black holes (for a review see {\it e.g.} \brmi),
%but is less known for BPS domain walls.

In this work we treat both systems in a unified framework
of the effective supersymmetric quantum mechanics with
a certain superpotential,
so that critical points of the superpotential correspond to
supersymmetric (anti de Sitter) vacua in the supergravity theory.
In the case of domain walls
the superpotential is inherited from the original supergravity
theory, and in the case of black holes it has the meaning
of the central charge of a black hole.
Notice, that in both cases physics requires us to extremize
these quantities.
Although the presence of the superpotential is crucial,
a lot of interesting questions can be answered without
precise knowledge of its form, simply assuming that it
is generic enough\foot{Later we will explain the precise
meaning of these words.}.

For example, domain walls (and similarly black holes) interpolating
between different minima have interpretation of the RG-flow in
the holographic dual field theory on the boundary. Therefore,
physically interesting questions about supersymmetric vacua and
RG-flow trajectories translate into classification of ground
states and gradient flows in the effective quantum mechanics.
Following the seminal work of Witten \witten,
we use the relation between supersymmetric quantum
mechanics and Morse theory to classify these
supersymmetric vacuum configurations.
Namely, according to Morse theory, the complex of critical
points with maps induced by gradient flow trajectories
is equivalent to the Hodge-de Rham complex of the scalar field manifold.
Notice, that the result is completely determined by the topology
of the scalar field manifold, but not the form of the superpotential.

With these goals and motivation, we start in the next section with the
derivation of supersymmetric quantum mechanics from five-dimensional
BPS domain walls which interpolate between different (AdS) vacua. In
the dual four-dimensional theory these solutions can be interpreted as
RG trajectories in the space of coupling constants.  We show that
these trajectories are gradient flows with a potential, so-called
height function, given by (the logarithm of) the superpotential in
supergravity theory.  We also explain the relation between critical
points of the height function and supersymmetric vacua of the
five-dimensional supergravity.  In section 3 we extend these results
to four-dimensional gauged supergravity, and also include interaction
with hypermultiplets which has not been studied until recently.  In
section 4 we draw a parallel with the black hole physics and, in
particular, show how quantum mechanics of the same type appears from
the radial evolution.  In this case, the height function is given by
the central charge of the black hole.  We also derive effective
potentials associated with membranes wrapped over holomorphic curves
in Calabi-Yau compactifications of M theory. After all these systems
are reduced to supersymmetric quantum mechanics, one might hope
to achieve classification of the critical points of the height function by means
of Morse theory, {\it cf.} \witten. We review the relevant topology and
explain its physical interpretation in section 5.  Finally, we put all
the ideas together in section 6 and demonstrate them in a family of
simple examples based on $SL(3)$ symmetric coset spaces.

%%%%%%%%%%%%%%%%%%%%%%%%%%%%%%%%%%%%%%%%%%%%%%%%%%%%%%%

\newsec{Quantum Mechanics of Domain Walls}

%%%%%%%%%%%%%%%%%%%%%%%%%%%%%%%%%%%%%%%%%%%%%%%%%%%%%%%%

BPS domain walls are kink solutions where the scalar
fields interpolate between different extrema of the supergravity
potential and due to the AdS/CFT correspondence, these solutions are
expected to encode the RG flow of the dual field theory.  Let us focus
here on the 5-d case with real scalars and postpone the modifications
for complex or quaternionic scalars for the next section.  A
Poincar\'e-invariant ansatz for the metric reads as follows:
\eqn\dwmetric{
ds^2 = e^{2U} \Big( -dt^2 + d\vec x^2 \Big) + dy^2 \ .
}
The function $U = U(y)$ is fixed by the equations of motion
coming from the variation of the action:
\eqn\action{
S = \int_M \Big( {R \over 2} - {1 \over 2} g_{ij} \partial \phi^i
\partial \phi^j - V \Big) \ - \ \int_{\partial M} K \ .
}
We included a surface term $K$ as the outer curvature which
cancels the surface contribution from the variation of the Ricci
scalar. There are no surface terms including the scalars because they
asymptotically extremize the superpotential and hence are constant
at the boundary. The form of the potential
\eqn\potential{
V = 6 \, \Big( \; {3 \over 4} \,
 g^{ij} \partial_i W \partial_j W - W^2 \; \Big) \
}
as a function of the superpotential $W$ is universal for a given
dimension
and follows from very general stability arguments \refs{\town}.
In fact, for complex and quaternionic scalar field manifolds it is also
possible to define a real superpotential $W$.  In this case the
supergravity
potential $V$ will have the similar form  \refs{\behe}; see the next
section.

The Poincar\'e invariance of the ansatz \dwmetric\ implies that all
worldvolume directions are Abelian isometries, so that we can integrate
them out.  For our ansatz the Ricci scalar takes the form
$R= - 20 (\dot U)^2 - 8 \ddot U$
and after a Wick rotation to an Euclidean time we
find the resulting 1-dimensional action\foot{In our notation,
dotted quantities always refer to $y$-derivatives.}:
\eqn\onedime{
S \sim \int \, dy \, e^{4U} \Big[ - 6 \, \dot U^2 +
{1 \over 2} g_{ij} \dot \phi^i \dot \phi^j + V\Big] \ .
% 4 \int dy {d \over dy}\Big(e^{4U} U' \Big)  \ .
}
In deriving this expression, the surface term in \action\ was canceled
by the total derivative term.  The equations of motion of this action
describe trajectories $\phi^i= \phi^i(y)$ of particles in the target
space $\CM$ with the metric $g_{ij}$. As a consequence of the 5-d
Einstein equations, these trajectories are subject to the constraint
\eqn\constr{
- 6 \, \dot U^2 + {1 \over 2} |\dot \phi^i|^2 - V = 0
}
with $|\dot \phi^i|^2 = g_{ij} \partial_y \phi^i \partial_y \phi^j$.
In order to derive the Bogomol'nyi bound
we can insert the potential into \onedime\ and write the action as
\eqn\bogom{
S \sim  \int \, dy \, e^{4U} \Big[ - 6 \, (\dot U \mp W)^2 +
{1 \over 2} \big| \dot \phi^i  \pm 3 \, \partial^i W \big|^2 \Big]
\mp 3  \int dy {d \over dy}\Big[e^{4U} \, W \Big]
}
leading to the BPS equations for the function $U=U(y)$ and
$\phi^i = \phi^i(y)$:
\eqn\bps{
\dot U = \pm W \qquad , \qquad
\dot \phi^i = \mp 3 \, g^{ij} {\partial W \over \partial \phi^j} \ .
}
If these equations are satisfied, the bulk part of the action
vanishes and only the surface term contributes. In the asymptotically
$AdS_5$ vacuum this surface term diverges near the AdS boundary
($U \sim y \rightarrow \infty $) and after subtracting the divergent
vacuum
energy
one obtains the expected result that the energy (tension) of the wall
is proportional to $\Delta W_0= W_{+\infty} - W_{-\infty}$.
%
%
% In this (extreme) case, the Bogomol'nyi bound is saturated, but if we
% consider instead a deSitter worldvolume, related to a non-vanishing
% cosmological constant on the world volume field theory, we get an
% additional contribution proportional to the cosmological constant. It
% does not enter directly the equations of motion for the particle
% trajectory $\phi^i(y)$, but the rhs in the constraint
% \constr\ becomes non-trivial and the value of the action is
% shifted (the Bogomol'nyi bound is not saturated anymore).

Our metric ansatz \dwmetric\ was motivated by Poincar\'e invariance
which is not spoiled by a reparameterization of the radial
coordinate. We have set $g_{yy}=1$, which is one possibility to fix
this residual symmetry. On the other hand, we can also use this
symmetry to solve the first BPS equation $W\, dy = \pm dU$, i.e.\
take $U$ as the new radial coordinate. In this coordinate system
the metric reads
\eqn\newmetric{
ds^2 = e^{2U} \Big(-dt^2 + d\vec x^2 \Big) + {dU^2 \over W^2} \ .
}
Repeating the same steps as before we obtain the Bogomol'nyi
equations for the scalars
\eqn\newbogom{ -  \dot \phi^i = g^{ij} \partial_j \log |W|^{3}
=  g^{ij} \partial_j h }
which follow from the one-dimensional action
\eqn\morseaction{ S \sim \int dy \, \Big[ \,
|\dot \phi^i |^2 + g^{ij} \partial_i h \partial_j h \Big]
= \int dy \, | \dot \phi^i +  g^{ij} \partial_j h|^2
+ \ ({\rm surface \ term})
}
where $h = 3 \log |W|$. As before, the field equations are subject to
the constraint ${1 \over 2} |\dot \phi^i|^2 - g^{ij} \partial_i h
\partial_j h = 0$ and the surface term yields the central charge.
Supersymmetric vacua are given by the extrema of $h$ and the number
and type of such vacua can possibly be determined by using Morse theory
where $h$ is called the height function, see \refs{\witten} and below.

If there are more than two smoothly connected extrema of $h$, we can
build kink solutions corresponding to domain walls in the 4- or
5-dimensional supergravity. Let us summarize the different types of
domain walls and discuss their implications for the RG flow and
Randall-Sundrum scenario, see also \refs{\cvgr}. As long as $W \neq 0$
at the extremum, we obtain an AdS vacuum and since the extrema of $W$
are universal (independent of the radial parameterization), we have to
reach the AdS vacuum either near the boundary ($U \rightarrow
+\infty$) or near the Killing horizon ($U \rightarrow -\infty$).
Obviously, the case $U \rightarrow + \infty$ corresponds to a large
supergravity length scale and therefore, due to the AdS/CFT
correspondence, describes the UV region of the dual field theory. The
opposite happens for $U \rightarrow - \infty$, which is related to
small supergravity length scales and thus encodes the IR behavior of
the dual field theory. Moreover, extrema of $W$ are fixed points of
the scalar flow equations and translate into fixed points of the RG
flow, which can be either UV or IR attractive. The universality of the
fixed points of the RG flow, e.g.\  the scheme independence of scaling
dimensions, translates in supergravity to the fact that
the properties of the extrema of the superpotential are independent
of the chosen parameterization of the scalar manifold.
%  e.g.\ the eigenvalues of the mass matrix are invariant under
%  the diffeomorphism $\phi \rightarrow \tilde \phi(\phi)$.

In order to identify the different fixed points we do not need to
solve the equations explicitly; as we will see, they are determined by
the eigenvalues of the Hessian of the height function $h$.
This data depends only on the local behavior
of the superpotential near the fixed point.
Let us go back to the BPS equations \bps\ and
expand these equations around a given fixed point with
$\partial_i W\big|_0 =0$ at $\phi^i = \phi^i_0$. The superpotential
becomes
$$
W = W_0 + { 1 \over 2} (\partial_i \partial_i W)_0 \, \delta \phi^i
\delta
\phi^j \pm \ldots
$$
with $\delta \phi^i = \phi^i - \phi^i_0$, and the cosmological constant
(inverse AdS radius) is given by
$\Lambda = - W_0^2 = -1/R_{AdS}^2$. Hence, the
scalar flow equations can be approximated by\foot{For definiteness
we took the upper sign convention.}:
\eqn\approx{
\delta \dot \phi^i = - (g^{ij} \partial_j \partial_k W)_0 \, \delta
\phi^k
\ .
}
Next, we can diagonalize the constant matrix $(g^{ij} \partial_j
\partial_k W)_0$ and find
\eqn\hessian{
 \Omega^i_{\ k} = (g^{ij} \partial_j \partial_k W)_0
= W_0 \, {1 \over 3} \, \Delta^{(i)}  \, \delta^i_{\ k}
}
where we absorbed the inverse length dimension into $W_0$. The
dimensionless
eigenvalues $\Delta^{(i)}$ coincide with the eigenvalues of
$\partial^i \partial_j h$. According to the AdS/CFT correspondence
\refs{\ADS}, these eigenvalues are the scaling dimensions of the
corresponding
perturbations in the dual field theory. Namely, in a linearized version,
the
equations of motion for the scalars become $\partial^2 \phi^i - M^i_{\
j} \phi^j=0$ and the mass matrix reads $M^i_{\ j} =
\partial^i\partial_j V\big|_0 = W_0^2 \, \Delta^{(i)} (\Delta^{(i)} -
4 ) \delta^i_{\ j}$ or, measured in the units of $W_0$, the
mass formula becomes
\eqn\massformel{
(m^{(i)})^2 = \Delta^{(i)} (\Delta^{(i)} - 4 ) \ .
}
Consequently, near the AdS vacuum we find a solution to \approx\
\eqn\scalar{
U = (y-y_0) W_0 \quad , \quad
\delta \phi^i = e^{- {1 \over 3} \, \Delta^{(i)} W_0 (y - y_0)} =
e^{- {1 \over 3}\, \Delta^{(i)} U} \ .
}
This approximate solution is, of course, valid only if
$\delta \phi^i = \phi^i -
\phi^i_0 \rightarrow 0$ in the AdS vacuum where $U \rightarrow \pm
\infty$
and therefore {\it all} eigenvalues $\Delta^{(i)}$ have the same
sign: $\Delta^{(i)} >0$ for UV fixed points ($U \rightarrow + \infty$),
or $\Delta^{(i)} <0$ for IR fixed points ($U \rightarrow -\infty$).
Equivalently, UV fixed points are minima of the height function $h$
whereas IR fixed points are maxima. For this conclusion
we assumed that the scalar metric has Euclidean signature and $W_0>0$.
It is important to notice that in the definition of the scaling
dimensions the matrix $\Omega^i_{\ j}$ has one upper index and one lower
index. It is straightforward to consider also the possibilities
$W_0 <0$ and/or timelike components of the scalar field metric.  Note,
the
sign ambiguity in the BPS equations \bps\ interchanges both sides of
the wall, i.e.\ it is related to the parity transformation $y
\leftrightarrow -y$, which also flips the fermionic
projector onto the opposite chirality.

The eigenvalues of $\Omega^i_{\ k}$ of different signs
mean that the extremum is
IR-attractive for some scalars and UV-attractive for the other and,
therefore, is not stable (a saddle point of $h$). Moreover,  using these
saddle points to connect two maxima/minima of $h$ would violate the
proposed $c$-theorem for domain walls \refs{\algo, \gipe, \frgu,
\be}. Namely, multiplying eq.\ \newbogom\ by $g_{ik} \dot \phi^k$ one
obtains
\eqn\ctheor{
- \dot h = - \dot \phi^i \partial_i h =
g_{ij} \dot \phi^i \dot \phi^j \geq 0 \ .
}
Therefore, along the flow, the height function $h$ has to behave
strictly monotonic and at the extrema it corresponds to the central
charge of the dual conformal field theory \refs{\hesk}: $c_{CFT} \sim
R_{ADS}^3 = 1/|W_0|^3 = e^{-h_0}$. Recall that in our sign convention
larger values of the radial parameter $U$ correspond to the UV region
and are minima of the height function $h$. If we start with the UV
point ($U = + \infty$) and go towards lower values of $U$, the
$c$-theorem states that $h$ has to increase, either towards an IR
fixed point (maximum) or towards a positive pole in $h$ ($W^2
\rightarrow \infty$), which is singular in supergravity and
corresponds to $c_{CFT} = 0$.  On the other hand, if we start from an
IR fixed point ($U=-\infty$) and go towards larger values of $U$, due
to the $c$-theorem $h$ has to decrease, either towards a minimum (UV
fixed point) or towards a negative pole ($W^2 \rightarrow 0$), which
is {\it not} singular in supergravity and corresponds to flat
spacetime, $c_{CFT} \rightarrow \infty$.
An example is the asymptotically flat 3-brane, where the height function
parameterizes the radius of the sphere, which diverges asymptotically
(indicating decompactification) and runs towards a finite value near
the horizon which is IR attractive in our language.

In summary, there are the following distinct types of supergravity
flows, which are classified by the type of the extremum of the height
function or superpotential. Depending on the eigenvalues of the
Hessian of $h$, the extrema can be IR attractive (negative
eigenvalues), UV attractive (positive eigenvalues) or flat space
(singular eigenvalues). Generalizing the above discussion and allowing
also possible sign changes in $W$, the following kink solutions are
possible (in analogy to the situation in four dimensions
\refs{\cvgr}):

$(i)$ flat $\leftrightarrow$ IR

$(ii)$ IR  $\leftrightarrow$ IR

$(iii)$ IR  $\leftrightarrow$ UV

$(iv)$ UV  $\leftrightarrow$ UV \qquad (singular wall)

$(v)$ UV $\leftrightarrow$ singularity \quad ($W^2 = \infty$).

Note, there is no kink solution between a UV fixed point and flat
space, because the $c$-theorem requires a monotonic $h$-function and
the UV point corresponds to a minimum of $h,$ whereas the flat space
case is a negative pole. Moreover, if there are two fixed points of
the same type on each side of the wall, $W$ necessarily has to change
sign implying that the wall is either singular (pole in $W$) or one
has to pass a zero of $W$. In addition, between equal fixed points no
flow is possible (that would violate the $c$-theorem) and therefore
this case describes a static configuration, where the scalars do not
flow. This is also what we would expect in field theory, where the
RG-flows go always between different fixed points.  Recall, although a
zero of $W$ means a singularity in $h$, the domain wall solution can
nevertheless be smooth.  In models with mass deformations
($\Delta^{(i)} = 2$), type (v) walls appear generically for models
which can be embedded into maximal supersymmetric models, whereas
models allowing type (iv) walls typically can not be embedded into
maximal supersymmetric models\foot{Maximal supersymmetric models
typically have only one UV extremum.}, an explicit example is
discussed in \refs{\becve}.

In the Randall-Sundrum scenario one is interested in the case (ii),
because in this case the warp factor of the metric vanishes
exponentially on each side of the wall.

%%%%%%%%%%%%%%%%%%%%%%%%%%%%%%%%%%%%%%%%%%%%%%%%%%%%%%%%

\newsec{Potentials from Gauged Supergravity}

%%%%%%%%%%%%%%%%%%%%%%%%%%%%%%%%%%%%%%%%%%%%%%%%%%%%%%%%%

Extrema of the supergravity potential $V$ are vacua of the theory, but
not all extrema correspond to stable vacua. Instead, one can show
\refs{\town} that stable vacua are extrema of a superpotential $W$
which defines the supergravity potential in $d$ dimensions to be
$$
V = {(d-2)(d-1) \over 2} \, \Big( \; {d-2 \over d-1} \,
 g^{ij} \partial_i W \partial_j W - W^2 \; \Big)  \ .
$$
Once again, the corresponding scalar flow equations look like $\dot
\phi^i
= - g^{ij} \partial_j h$ with the height function
$$
h = (d-2) \log|W|.
$$
We will now derive the real superpotential $W$ for different
models. Depending on the number of unbroken supercharges, only special
superpotentials can appear in supersymmetric models. Our primary
interest are supergravity duals of field theories with 4 unbroken
supercharges. Therefore the scalars on the supergravity side are part
of vector, tensor or hyper multiplets that parameterize the product
space:
$$
\CM = \CM_{V/T} \times \CM_H.
$$
Known potentials
are related to: (i) gauging isometries of $\CM$ or (ii) gauging the
global
R-symmetry. In the first case, the scalars and fermions become charged
whereas in the second case the scalars remain neutral.  As a
consequence of the gauging, the supersymmetry variations are altered
and flat space is, in general, not a consistent vacuum. If the potential
has an extremum, it is replaced by an AdS vacuum, which was not
welcomed in the early days of supergravity, but fits very well in the
AdS/CFT correspondence and the holographic RG flow picture.  Let us
discuss the different cases in more detail.

%%%%%%%%%%%%%%%%%%%%%%%%%%%%%%%%%%%%%%%%%%%%%%%%%%

\subsec{Gauged Supergravity in 5 Dimensions}

%%%%%%%%%%%%%%%%%%%%%%%%%%%%%%%%%%%%%%%%%%%%%%%%

Supergravity in five dimensions needs at least 8 supercharges, and
scalar fields can be part of vector-, tensor- or hypermultiplets.  Each
(abelian) vector multiplet contains a $U(1)$ vector $A_{\mu}^i$, a
gaugino $\lambda^i$ and a real scalar $\phi^i$ ($i=1, \ldots ,
n_V$). On the other hand, a hypermultiplet includes two hyperinos
$\zeta^u$ and four real scalars $q^u$ ($u = 1, \ldots ,
4n_H$). Finally, the gravity multiplet has besides the graviton,
the gravitino $\psi^{A}_m$ and the graviphoton $A_{\mu}^0$.

%%%%%%%%%%%%%%%% changed part %%%%%%%%%%%%%%%%%%%

On the vector multiplet side, supersymmetry is a powerful tool in
determining allowed corrections. In fact, all couplings entering the
Langrangian are fixed in terms of the cubic form \refs{\gusii}
\eqn\Fconstr{
F = {1 \over 6} C_{IJK} X^I X^J X^K \ .
}
In a Calabi-Yau
compactification of M-theory, the fields $X^I$ ($I = 0, \ldots , n_V$)
are related to the K\"ahler class moduli $M^I$ by a rescalling
$X^I={M^I \over {\cal{V}}^{1/3} } $ with the Calabi-Yau volume
${\cal{V}}={1\over 6} C_{IJK} M^I M^J M^K$
and the constants $C_{IJK}$ are the topological
intersection numbers \refs{\cace}. The scalar fields $\phi^a(X^I)$
parameterize the space $\CM_V$ defined by $F=1$ and the gauge- and
scalar-couplings are given by
\eqn\MXmetr{
G_{IJ} = -{1\over 2}(\partial_I \partial_J
F)_{F=1}, \quad \,\, g_{ij} = (\partial_i X^I \partial_j X^J
G_{IJ})_{F=1}.}

Much less is known on the hypermultiplet side.  The four real scalars
of each hypermultiplet are combined to a quaternion and parameterize a
quaternionic mannifold $\CM_H$.  {From} the geometrical point of view,
the hypermultiplet sector in four and five dimensional supergravity is
the same, see \refs{\anbe} for further details.  In any
compactification from string or M-theory there is at least one
hypermultiplet -- the so-called universal hypermultiplet that contains
Calabi-Yau volume ${\cal{V}}.$ This name is a little bit misleading
since when $n_H > 1$ there is no unique way to single out one
direction on a general quaternionic manifold \Aspinwall.  As long as
we have only this single hypermultiplet, its geometry is given by the
coset space ${SU(2,1) \over U(2)}$. But after including further
hypermultiplets, quantum corrections will deform the space in a way
which is rarely known.

Now let us turn to potentials resulting from gauged isometries.  Both
manifolds, $\CM_V$ and $\CM_H$, have a number of isometries
\refs{\wipr}, which can be gauged \refs{\guza, \ceda}.
However the flow equations of the scalars are not sensible to gauged
isometries of $\CM_V$; it yields only an additional ``D-flatness''
constraint\foot{See also below for the similar situation in 4
dimensions.}
\refs{\kali, \ceda}. More interesting is the gauging of
isometries of $\CM_H$; see
\refs{\luov,\behe,\becvd} for explicit examples.
This is a quaternionic space, which implies
the existence of three complex structures and an associated triplet of
K\"ahler forms $K^x$. The holonomy group is $SU(2) \times Sp(n_H)$
and the K\"ahler forms have to be covariantly constant with
respect to the $SU(2)$ connection. The isometries are
generated by a set of Killing vectors $k_I^u$
$$ q^u \rightarrow q^u + k^u_I \epsilon^I $$
and the required gauge covariant derivatives become $dq^u \rightarrow
dq^u + k_I^u A^I$. In order to keep supersymmetry, this gauging has to
preserve the quaternionic structure, which means that the Killing
vector has to be tri-holomorphic (in analogy with the holomorphicity
in $N$=1 supergravity). This is the case if we can
express them in terms of a triplet of Killing prepotentials $P^x_I$
(with the $SU(2)$ index $x =1,2,3$)
\eqn\killing{
K^x_{uv} k^v_I = - \nabla_u P_I^x \equiv - \partial_u P_I^x -
\epsilon^{xyz} \omega^y_u P^z_I \ .
}
Here $\omega_u^y$ are the $SU(2)$ connections related to the
K\"ahler forms by $K^x_{uv} = - \nabla_{[u} \omega^x_{v]}$. They can
be also expressed in terms of the complex structures and the
quaternionic metric as: $K^x_{uv} = (J^x)_{u}^{\ r} h_{rv}$, and
using $\sum_x (J^x)_u^{\ r} (J^x)_r^{\ v} = - 3 \, \delta_u^{\ v}$
we can write the Killing vectors as $k^u_I = - \sum_x h^{uv}
(J^x)_v^{\ r} \nabla_r P_I^x$. Next, introducing an $SU(2)$-valued
superpotential
\eqn\superpot{
W_A^{\ B} \equiv  W^x \, (i \sigma_x)_A^{\ B}
 \qquad {\rm with:} \quad W^x= X^I P_I^x  =
 X^I(\phi)\, P_I^x(q)
}
($\sigma_x$ are the Pauli matrixes) the fermionic supersymmetry
variations \refs{\ceda} become
\eqn\killingeqs{\eqalign{
\delta\psi_m^{A}  = & D_m\epsilon^{A} - {i \over 3}\,W_B^{\ A}\,
 \Gamma_m\epsilon^B \ , \cr
\delta \lambda^{Ai} =& -{i \over 2} \Big[ \Gamma^m \partial_m \phi^i\,
 \epsilon^{A} - 2 i\,   g^{ij} \partial_j W_B^{\ A}\, \epsilon^B \,
 \Big]  \ , \cr
\delta\zeta^\alpha =&  -{i \over \sqrt{2}}\, V^{A\alpha}_u\,\Big[
 \Gamma^m \partial_m q^u\, - 2  h^{uv} (J^x)_v^{\ r}
 \nabla_r W^x) \Big] \epsilon_A
}
}
where $A,B$ are $SU(2)$ indices and $\partial_i \equiv {\partial
\over \partial \phi^i}$. We dropped the gauge field contributions,
since they are not important for {\it flat} domain wall solutions;
they would be important e.g.\ for a worldvolume geometry $R \times
S_3$.  For supersymmetric vacua, $W_A^{\ B}$ has to become extremal
($\partial_i W^x = \nabla_u W^x = 0$, for all $x = 1,2,3$). One can
show \refs{\behe}, that the $SU(2)$ phase of $W$ does not contribute
to the scalar flow equations and is absorbed by the $SU(2)$ connection
entering the covariant derivative $D_m$. Moreover, combining all
scalars $(\phi^i , q^u)$ that parameterize the space $\CM =
\CM_V \times \CM_H$ with the metric $g_{ij} = {\rm diag}(g_{ij} ,
h_{uv})$
one recovers the BPS equation \bps\ with the real-valued
superpotential
\eqn\rsuperp{
W^2 = \sum_x W^x W^x \ .
}
There is one especially simple example, where the superpotential is
only $U(1)$ valued, i.e.\ the Killing prepotential has only one, say,
$P^3_I$ component. In this case the $SU(2)$ covariant derivative in
\killing\ becomes a partial derivative and we have the
freedom to shift the Killing prepotential by any constant $P^3_I
\rightarrow P^3_I + \alpha_I$. In fact, one can even set the Killing
prepotential to zero and keep only the constants $\alpha_I$, which are
the analogs of the FI-terms in field theory.  As a consequence, the
Killing vectors vanish as well as the charges of the scalars.
But still, we have a non-trivial potential giving a mass to all vector
scalars. In this case the superpotential becomes
\eqn\superpotf{
W^{(3)}(\phi^i) = (\alpha_I X^I)_{F=1}\ ,
}
which is manifestly real valued. Due to the constraint $F=1$ this
potential yields an AdS vacuum, where generically all vector-scalars
are fixed and the moduli space $\CM_V$ of vacua is lifted
except for a discrete set of extremal points of $W^{(3)}$.
{From} the 5-d supergravity
perspective this model has been discussed in \refs{\gusi},
and important for the RG-flow is the property \refs{\gusii, \kama}
\eqn\secondder{
\partial_i \partial_j W^{(3)} = {2 \over 3} g_{ij} W^{(3)} +
 T_{ijk} \partial^k W^{(3)} \ .
}
This relation implies that all scaling dimensions, as defined in
\hessian, are $\Delta^{(i)} = +2$ and fulfill a sum rule
\refs{\khwa}: $\sum_i \Delta^{(i)} = 2 n$, with $n = {\rm dim}\CM_V$.
Therefore, the flow is generated by mass deformations in the field
theory and the positive sign indicates that all fixed points are UV
attractive; IR attractive critical points are excluded for this model
\refs{\kali, \becvv}.  This model can be obtained from Calabi-Yau
compactification of M-theory in the presence of non-trivial $G$-fluxes
parameterized by $\alpha_I$ \refs{\luov} and the superpotential can be
written as \refs{\gu, \begu}
\eqn\superpotn{
W^{(3)} = \int_{CY} K \wedge G_{flux}
}
where $K$ is the K\"ahler 2-form. However this compactification yields
an
un-stabilized Calabi-Yau volume and, as long as we treat it as a
dynamical field \refs{\tava, \begu}, this compactification does
not give flat space or AdS vacua. Nevertheless this run-away problem can
be
avoided by more general hypermultiplet gauges, see \refs{\behe}. It
would be interesting to derive the general $SU(2)$-valued
superpotential in the same way from M-theory. For a recent study of
$W^x$
(with $W^3 = 0$ but $W^1 \ne 0$ and $W^2 \ne 0$) see \CKLT.

%%%%%%%%%%%%%%%%%%%%%%%%%%%%%%%%%%%%%%%%%%%%%%%%%%%%%%%%%%%%%%%%

\subsec{Gauged Supergravity in 4 Dimensions}

%%%%%%%%%%%%%%%%%%%%%%%%%%%%%%%%%%%%%%%%%%%%%%%%%%%%%%%%%%%%%%%%

Supergravity in 4 dimensions needs at least 4 supercharges and allows
for more general (holomorphic) superpotentials which are not related
to gauged isometries. In the generic case, these models have an $AdS$
vacuum with a dual 3-d field theory with only 2 (unbroken)
supercharges; for domain wall solutions see \refs{\cvgr}.

However, if we again focus our attention on models with 8
supercharges, potentials have to be related to gauged isometries. The
main difference from the situation in 5 dimensions concerns the vector
multiplet side.  Since vector fields in four dimensions have only two
on-shell degrees of freedom, each vector multiplet has to contain two
scalars (in order to complete the bosonic degrees of freedom). These
two real scalars can be combined into a complex scalar $z^i$ and
supersymmetry requires that they parameterize a special K\"ahler
space, see \refs{\anbe} for a review.  On the other hand, the scalars
$q^u$ entering the hypermultiplets parameterize again a quaternionic
space and the gauging of the corresponding isometries goes completely
analogous to the case in 5 dimensions. This time, however, it is
reasonable to use the symplectic notation of special geometry and we
will use the holomorphic symplectic section ${\bf V} = \big(X^I(z) ,
F_I(z) \big)$ with the symplectic product defining the K\"ahler
potential:
\eqn\Kpot{
e^{-K/2} = \langle
\bar{ {\bf V}} , {\bf V}
\rangle = i( \bar X^I F_I - X^I \bar F_I).}
% Similarly, one defines symplectic
% gauge field strength ${\bf F}_{\mu\nu} = ( F^I_{\mu\nu} , G_{I\,
% \mu\nu})$ and one can decompose the Killing vectors of $\CM_V$ with
% respect to these symplectic gauge vector, i.e.\ ${\bf k}^i = (k^{i\,I}
% , k^i_{I})$ with $i=1, \ldots , n_V$. Then, we arrive at an analogous
% superpotential as in the 5-dimensional case (see eq.\ \superpot\ )
%
%
% \eqn\superpotfour{\eqalign{
% W_A^{\ B} \equiv  W^x \, (i \sigma_x)_A^{\ B}& =
%  [X^I \, P_I^x  - F_I \, P^{I\, x} ]\, ( i \sigma_x)_A^{\ B} \cr & =
%  \Big[X^I(z) \, P_I^x(q)  - F_I(z) \, P^{I\, x}(q)\Big]\,
%  (i \sigma_x)_A^{\ B}
% }
% }
Then, we obtain a similar superpotential, {\it cf.} eq.\ \superpot :
\eqn\superpotfour{
W_A^{\ B} \equiv  W^x \, (i \sigma_x)_A^{\ B} \qquad {\rm with} \quad
 W^x = X^I \, P_I^x   =
  X^I(z) \, P_I^x(q) \ .
}
Notice, now $X^I= X^I(z)$ is a complex field.
In $N$=2 supergravity in 4 dimensions one also
introduces a symplectic vector
${\bf F}_{\mu\nu} = ( F^I_{\mu\nu} , G_{I\, \mu\nu})$ for the gauge
fields.
Both gauge fields are related to each other and
we took the freedom to transform all gauge fields to $F^I_{\mu\nu}$
and therefore only the $X^I$ component of the section enters $W$. As
in 5 dimensions we can express the supersymmetry variations
\refs{\wila, \anbe} in terms of the superpotential \superpotfour
\eqn\killingeqsfour{\eqalign{
\delta\psi_m^{A}  = & D_m\epsilon^{A} - {i \over 3}\, e^{K/2}\, W_B^{\
A}\,
 \gamma_m\epsilon^B \ , \cr
\delta \lambda^{Ai} =& -{i \over 2} \Big[ \gamma^m \partial_m z^i\,
 \epsilon^{A} + 2 i \, e^{K/2}\,  g^{i \bar{j}} \nabla_{\bar j}
 \bar{W}_B^{\ A}\, \epsilon^B \,  + k_I^i \bar{X}^I e^{K/2}
 \epsilon^A \, \Big]  \ , \cr
\delta\zeta^\alpha =&  -{i \over \sqrt{2}}\, V^{A\alpha}_u\,\Big[
 \gamma^m \partial_m q^u\, - 2\, e^{K/2}\,  h^{uv} (J^x)_v^{\ r}
 \nabla_r W^x) \Big] \epsilon_A
}
}
with $\nabla_j W^x = P^x_I(q) \Big({\partial \over \partial z^j} +
{\partial K \over\partial z^j}\Big) X^I(z)$ as  a K\"ahler covariant
derivative and $\nabla_r W^x$ denotes the $SU(2)$ covariant derivative,
see \killing.  We included also a possible gauging of $\CM_V$ related to
the Killing vector $k_I^i$ and remarkably this gauging affects only
the gaugino variation $\delta \lambda^{Ai}$, but not the gravitino
variation $\delta \psi_{m}^A$.

As usual the fermionic projector is derived from the timelike
gravitino variation $\delta \psi_0^A$ and using the metric ansatz
\newmetric\ this projector becomes
\eqn\projector{
\gamma_U \epsilon^A \pm i \, {W_B^{\ A} \over |W|} \epsilon^B = 0
}
and we define a  real superpotential by
\eqn\foursp{
W^2 = e^{K} |W|^2 = \sum_x e^K W^x \bar W^x = \sum_x e^K X^I \bar X^J
 P^x_I P^x_J
}
(the hyper scalars $q^u$ are $4n_H$ real fields so that the
Killing prepotentials are real). Note, the projector \projector\
contains
both, the $SU(2)$ phase as well as the $U(1)$ phase related to the
complex fields $X^I$ and in order to ensure the vanishing of the
radial gravitino variation $\delta \psi_U^A$, both phases have to be
absorbed into the $SU(2)$ (resp.\ $U(1)$) K\"ahler connection.  This can
impose further constraints.

Using the projector it is straightforward to show that the
gaugino variation yields two equations\foot{In this matrix equation,
the coefficient in front of each Pauli matrix has to vanish. In
addition, in all these calculations one has to keep in mind that the
$\gamma$-matrix implicitly contains a $W$ factor due to its curved
index.}: a $D$-flatness constraint $k^i_I \bar X^I = 0$ and, after
fixing the sign ambiguity in the projector, the expected flow equation
for the scalars $z^i$:
\eqn\zequ{
\dot z^i = - g^{i \bar j} \partial_{\bar z^j} \log W^2 \ .
}
Here we use: $\partial_{\bar i} W^2 = 2 \, W \partial_{\bar
i} W = \sum_x e^K W^x (\partial_{\bar i} + \partial_{\bar i} K)\bar W^x
=
\sum_x {W^2 \over |W|^2} W^x \nabla_{\bar i} \bar W^x$.
Finally, following the steps done in 5 dimensions, see also
\refs{\behe}, the hyperino variation yields the same flow equation
$$\dot q^u = - h^{uv} \partial_v \log W^2 \ .$$

As before, we can again consider the special case, where the Killing
prepotentials have only one component (e.g.\ $P^1_I=P^2_I=0,\ P^3_I
\neq 0$) and can be shifted by arbitrary constants $\alpha_I$.
The analog of \superpotf\ is now the complex superpotential
\refs{\anbe}
$$
W^{(3)} = \alpha_I X^I(z) \ .
$$
Recall that this form is related to the special symplectic basis, where
all $G_{I\,\mu\nu}$ gauge fields have been dualized to $F^I_{\mu\nu}$
gauge fields. In general, the superpotential has the covariant form
\refs{\mi, \tava}
\eqn\superpotff{
W^{(3)} = \alpha_I X^I - \beta^I F_I \ .
}
It allows for AdS vacua which are in one-to-one correspondence with
solutions to the attractor equations, which determine extrema of the
supersymmetry central charge \refs{\feka, \fekaa}.  One can show that
all these extrema are UV attractive \refs{\becvv} (due to a similar
relation as \secondder) and, therefore, cannot give regular domain wall
solutions.  Like in 5 dimensions, this superpotential is related
to a compactification in the presence of fluxes. This time, however, it
is type IIB Calabi-Yau compactification and the superpotential
can be written as \refs{\tava, \gu}
$$
W^{(3)} = \int_{CY} \Omega \wedge H_{flux}
$$
where $\Omega$ is the holomorphic (3,0)-form and
$H_{flux} = H^{RR}_{flux} + \tau H^{NS}_{flux}$
with $\tau$ as the complexified scalar field of type IIB string theory.
%Due to the un-stabilized dilaton, such a
%compactification with background flux yields a run-away potential, as
%in 5 dimensions.
%
% and there is no AdS or flat space vacuum, although
% one may consider certain limits where this mode decouples \tava.
Supersymmetric vacua corresponding to the minima of
this superpotential have been recently studied in \CKLT.
Moreover, the authors of \CKLT\ pointed out a relation
between these supersymmetric vacua and attractor points
of $\CN=2$ supersymmetric black holes.
Namely, they demonstrated the equivalence of the supersymmetry
conditions $\nabla_i W^x = 0$ to the attractor equations
that determine the values of the scalar fields $z^i$ at the horizon
of a black hole. We elaborate this relation in the next section
from the standpoint of the effective quantum mechanics,
whereas in the rest of this section we discuss `attractor'
interpretation of the other supersymmetry condition:
\eqn\hypersusy{\nabla_u W^x =
X^I \Big( \partial_u  P^x_I  + \epsilon^{xyz} \omega_u^y P^z_I \Big) = 0
\ .}

The idea is that the supersymmetry variations of the hyperino
\killingeqs\ (or \killingeqsfour) describe supersymmetric
vacua of $\CN=2$ gauged supergravity as well as
BPS instantons (in ungauged supergravity)
responsible for non-perturbative corrections to the metric
on the moduli space of hypermultiplets $\CM_H$.
In order to see the relation more precisely, consider
supergravity theory obtained from Calabi-Yau compactification
of M theory (or type IIA string theory).
Then, the instantons in questions are constructed from
(bound states of membranes and)
Euclidean 5-branes wrapped on the Calabi-Yau manifold \GS.
Put differently, this BPS configuration can be understood
as a five-brane wrapped around the Calabi-Yau space with
world-volume 3-form tensor field turned on.

Apart from the number of 5-branes, the instanton is characterized
by the membrane charge $\alpha_u$ that takes values in
the homology lattice $H_3 (CY, \Z)$.
For a given set of the charges, one can construct
a spherically symmetric solution $q^u (r)$ which
preserves half the supersymmetry \killingeqs\ and
behaves like\foot{For the sake of concretness, we consider
five-dimensional case obtained from compactification of M theory.
Reduction to four dimensions is straightforward.} \refs{\GS,\Sabra}:
\eqn\asympt{q^u (r) \sim {\alpha_u \over 3 r^3} + {\rm const},
\quad r \to \infty \ .}
When $r$ varies from $r=\infty$ to $r=0$, the scalar fields $q^u$
go to the fixed values determined by the charges $\alpha_u$,
similar to the attractor mechanism \feka. It was shown in \GS,
that the radial evolution of the scalar fields $q^u (r)$
is described by null-geodesics in the effective 0+1 dimensional
theory with target space $\CM_H$. Up to boundary terms, we expect
the action of this effective theory is equivalent to
the action of supersymmetric quantum mechanics \morseaction\
with the superpotential \superpot\ (or \superpotfour).
Thus, we claim that the flow on the moduli space of
the hypermultiplets is governed by $W^x$.

Since the instanton solution is expected to be smooth
at $r=0$, from the equation \killingeqs\
(with the $U(1)_R$ gauge field contribution included)
we find that the fixed points of the flow on the moduli space
of hypermultiplets are characterized by the condition \hypersusy:
\eqn\hyperattr{X^I k_I^u = 0 \ .}
This simple condition means that the Killing vectors obey certain
linear relations at the fixed points in the space $\CM_H$.
Note, that in the hypermultiplet version of the attractor
mechanism we have fixed limit cycles rather than fixed
points\foot{We would like to thank A.~Strominger for pointing
this out to us and drawing our attention to the hypermultiplet
version of the attractor mechanism as we describe it here.}.
It would be interesting to better understand  the physics
and the geometry of these fixed cycles in $\CM_H$ \progress.

Intuitively, the hypermultiplet attractor equation \hyperattr\
could be expected by analogy with $\CN=2$ BPS black hole
solutions that exhibit enhancement of supersymmetry at the horizon.
For a similar reason, one might expect enhancement of
supersymmetry at the center, $r=0$, of the BPS instanton
constructed from a bound state of a membrane wrapped around
a special Lagrangian cycle and a five-brane wrapped on
the entire Calabi-Yau space.
As we explained in the earlier sections, a supersymmetric vacuum
is given by the extremum of the superpotential \hypersusy\
which, in turn, is equivalent to \hyperattr.

%%%%%%%%%%%%%%%%%%%%%%%%%%%%%%%%%%%%%%%%%%%%%%%%%%%%%%%%

\newsec{Black Hole -- Domain Wall Correspondence}

Similarity between physics of black holes of ungauged supergravity and
the domain wall solutions in gauged supergravity with gauging of
vector multiplets \refs{\kls, \bhk} is based on the fact that both
theories can be described by the same one-dimensional effective
Lagrangian. In both cases the solution is fixed by the same set of
attractor equations \refs{\feka, \fekaa} and the superpotential $W$ of
gauged supergravity corresponds to the supersymmetry central charge $Z$.

The D=4 N=2 action of ungauged supergravity in terms of special
geometry has a well known form:
\eqn\bhaction{
{\cal L} = {{R}\over{2}}-G_{i \bar j}\partial^{\mu} z^i\partial_{\mu}
\bar z^{\bar j} -h_{uv}\partial^{\mu} q^u\partial_{\mu} q^v - {\rm
Im}{\cal N}_{\Lambda\Sigma}{\cal F}^{\Lambda}{\cal F}^{\Sigma}-{\rm
Re}{\cal N}_{\Lambda\Sigma}{\cal F}^{\Lambda}{}^\ast{\cal F}^{\Sigma}
}
where $z$ as in Section (3.2) are the complex scalars of vector
multiplets parameterizing special K\"ahler manifold $\CM_V$ with
metric $G_{i \bar j}={{\partial^2K}\over{\partial z^i\partial\bar
z^{\bar j}}}$ and K\"ahler potential $K$. Their kinetic term can
be written in the form similar to the one in \action :
$g_{ij}d\phi^id\phi^j= G_{i \bar j}\partial z^i\partial \bar z^{\bar
j}.$  The vector couplings ${\rm Im}{\cal N}$ and ${\rm Re}{\cal N}$
depend only on scalar fields $z.$ Real scalars $q^u$ belong to
hypermultiplets and parameterize the quaternionic manifold $\CM_H$ with
the metric $h_{uv}$. In ungauged supergravity the hypermultiplets are
decoupled from the theory and we will consider only vector multiplet
scalars.  We will choose the following ansatz for D=4 extreme
black hole metric \FGK :
$$
ds^2 = -e^{2U}dt^2 + e^{-2U}\left({{d\tau^2}\over{\tau^4}}+{{1}
\over{\tau^2}}d\Omega_2^2\right).$$
This assumption leads to one-dimensional effective action of the
familiar form:

$$
S_{eff} \sim \int{\left(
\left({{dU}\over{d\tau}}\right)^2+g_{ij}{{d\phi^i}\over
{d\tau}}{{d\phi^j}\over{d\tau}}+e^{2U}V(\phi,p,q)\right)}.$$
 The potential $V$ depends on the symplectic covariant electric and
magnetic charges $(p^I , q_I)$ for symplectic vector ${\bf F}_{\mu\nu} =
(
F^I_{\mu\nu} , G_{I\,\mu\nu})$ and  can be identified with the
symplectic
invariant form $I_1$  in terms of the  complex central charge (the
graviphoton
charge) $Z = q_I X^I - p^I F_I$  with the symplectic section $(X^I ,
F_I)$:
\eqn\bhpotential{
V(\phi^i,p,q)=I_1=|Z(z,p,q)|^2+|\nabla_iZ(z,p,q)|^2
}
where $\nabla_i$ is a K\"ahler covariant derivative. Action \bhaction
$\,$
can be easily transformed to the standard form of Bogomol'nyi bound:
$$
{\cal L} = \left( {{dU}\over{d\tau}} \pm e^U|Z|\right)^2+{\big
|}{{dz}\over{d\tau}}\pm e^UG^{i\bar j}\bar \nabla_{\bar j}\bar Z{\big
|}^2 \pm
{{d}\over{d\tau }}\lbrack e^U Z\rbrack
$$
with the equations of motion:
\eqn\bheqm{
{{dU}\over{d\tau}} = \pm  e^U |Z|,\quad \,\,
{{dz^i}\over{d\tau}} = \pm e^U G^{i\bar j}\bar \nabla_{\bar j}\bar Z.
}
Using the same construction as in Section 3.2 for \zequ  $\,$ we finally
derived  the equation for the scalar field z:
$$
{{dz^i}\over{dU}}= G^{i\bar j}\partial_{\bar j}\log Z^2.
$$
This form is consistent with one-dimensional supersymmetric quantum
mechanics.

For the black hole solutions in D=5 ungauged supergravity 
\refs{\fekaa,\FGK,\Sabra} the situation is
similar. The action of D=5 N=2 ungauged supergravity coupled to vector
multiplets has a form:
$$
 S_5 =\int{ {R\over 2}-{1\over 4} G_{IJ}F^IF^J-{1\over 2}g_{ij}\partial
\phi^i\partial \phi^j -{e^{-1}\over
48}\epsilon^{\mu\nu\rho\sigma\lambda}
C_{IJK} F^J_{\mu\nu}F^J_{\rho\sigma}A^K_\lambda  }.
$$
Scalar fields $\phi^i$ are defined through the constraint \Fconstr,
$ F={{1}\over{6}} C_{IJK}X^IX^JX^K =1 $,
and the gauge coupling $G_{IJ}$ and $g_{ij}$ are given by \MXmetr\
and depend only on $X^I$ and $C_{IJK}$.
The  ansatz for the D=5 black hole solutions is:
\eqn\dfmet{
d s^2 =-e^{-4U}dt^2+e^{2U}\left(dr^2+r^2d\Omega_3^2\right),
}
and the ansatz vector fields is: $ G_{IJ}F^I_{tr}={1\over
4}\partial_rK_J$
where $ K_I = k_I+{Q_I \over r^2}$ and $Q^I$ are black hole electric
charges.

Using a new radial variable $\tau ={1 \over r^2}$ and equations of
motion it is
easy to reduce D=5 action to the effective D=1 dimensional action:
\eqn\dfactbh{
S_{eff} \sim \int{\left(
3\left({{dU}\over{d\tau}}\right)^2+{{1}\over{2}}
g_{ij}{{d\phi^i}\over{d\tau}}{{d\phi^j}\over{d\tau}}+e^{4U}{{1}\over{12}}V(\phi,p,q)\right)}}
where the potential $V$ comes from the kinetic term for the vector
fields
\eqn\bhpotfo{
V = {3\over 2}Q_IQ_JG^{IJ}=Z^2+{{3}\over{2}}g^{ij}\partial Z_i\partial Z_j}
where $Z=X^IQ_I$ is a central charge. In \bhpotfo\ we used special
geometry
relations (see for example \bhsusy):  $
g_{ij}=G_{IJ}X^I_{,i}X^J_{,j}=-3C_{IJ}X^I_{,i}X^J_{,j}$ and $
g^{ij}X^I_{,i}X^J_{,j}=G^{IJ}-{2\over 3}X^IX^J$ where
$X^I_{,i}={\partial X^I
\over \partial \phi^i}.$

 The effective action \bhpotfo\ is easily changed to the familiar BPS
form:
$$
{\cal L} = 3\left( {dU \over d\tau } \pm
{{1}\over{6}}e^{2U}Z\right)^2+{{1}\over{2}}{\big |}{ d\phi^i \over d\tau
}\pm
{{1}\over{2}} e^{2U} \partial^i Z {\big |}^2\mp
{{1}\over{2}}{{d}\over{d\tau
}}\lbrack e^{2U} Z\rbrack
$$
where $|{ d\phi^i \over d\tau }|^2=g_{ij}{d\phi^i \over d\tau}{ d\phi^j
\over
d\tau}.$  Once again we have a system of linear differential equations
of the
form:
$${\partial U\over \partial \tau} \pm {1\over 6}e^{2U} Z=0 ,\,\, \,
g_{ij}{\partial \phi^j \over \partial  \tau} \pm {1\over 2}e^{2U}
{\partial Z \over \partial\phi^i}=0 .$$
Using the first equation to introduce the new radial variable
$dU ={1\over 6}e^{2U}d\tau $ we get the familiar equation
for the gradient flow in a scalar field manifold $\CM_V$:
$${\partial \phi^i\over \partial U}= g^{ij} \partial_j\log Z^3 .$$

This consideration shows that both cases -- black holes of ungauged
supergravity coupled to vector multiplets and domain walls of gauged
supergravity can be treated using the same approach of one-dimensional
supersymmetric quantum mechanics.

The first interesting question connected to the above discussion is the form
of the black hole potential \bhpotfo\ .  The same potential appears as a result
of M-theory compactification on Calabi-Yau threefolds in the presence of
non-trivial G-fluxes  \refs{\luov, \begu ,\ceda} and corresponds to
gauging of the universal hypermultiplet of D=5 dimensional theory. 
An electrically charged five dimensional black hole corresponds
to a membrane in M-theory
wrapped around 2-cycles of Calabi-Yau manifold in the process of
compactification. We can formally regard the M-theory compactification
in the presence of the membrane-source and corresponding flux.

The 11-dimensional supergravity theory is described by the action:
$$
S_{11}={1\over 2}\int_{M^{11}}\left( \sqrt{-g} R - {1\over 2}G\wedge
* G -{1 \over 6}C \wedge G \wedge G  \right)
$$
 where $C$ is a 3-form field with 4-form field strength $G^{(4)}=dC$.
 In the presence of an electrically charged membrane source Bianchi
identities
and equations of motion for $D=11$ theory are:
\eqn\bioo{dG = 0}
\eqn\eqmooo{
d^{\star} G = d ^{\ast} G +{1\over 2}G\wedge G = 2k^2_{11} ( ^{\ast} J )}
where $^\star G = {\partial {\cal L}\over \partial G},$ $ ^{\ast} $ is
Hodge-dual and $J$ is a source current with  the corresponding Noether
``electric'' charge
$$
Q={\sqrt{2}k_{11}}\int_{ M_8} (^{\ast} J)={1\over \sqrt{2}k_{11}}\int_{
S_7}(^\star G)^{(7)}
$$
where $M_8$ is a volume and $S^7$ is a sphere around the membrane.
A formal solution to the equations of motion \eqmooo\ reads:
$$
^{\star} G = ^{\ast} G +{1\over 2}C\wedge G =
\sqrt{2}k_{11}Q{\epsilon_7\over
\Omega_7}
$$
where $\epsilon_7$ is a volume form and $\Omega_7$ is a 7-volume. This
solution
corresponds to the vector potential of the form:
\eqn\eqmsol{
C \sim {Q \over {\tilde r}^6}\omega_3 }
where ${\tilde r}$ is a $D=11$ transverse distance to the membrane and
$\omega_3$ is the  membrane volume form. Compactification from  $D=11$
dimensions down to $D=5$ leads to the configuration when the membrane is
wrapped around compact dimensions.

A natural splitting of moduli coordinates $M^a\,\Rightarrow \,
(X^I={M^I\over
{\cal{V}}^{1/3}};{\cal{V}})$ and the condition for function \Fconstr\ :
$F(X)=1$ defines $(h^{1,1}-1)$-dimensional hypersurface on
the Calabi-Yau cone and $(h^{1,1}-1)$ independent coordinates
$\phi^a(X^I)$ (special coordinates) on this  hypersurface define vector
multiplet moduli space with K$\ddot a$hler metric  $G_{IJ}$ \MXmetr\ :
 \eqn\vectmet{
G_{IJ}(X)= {i\over 2{\cal{V}}}\int \omega_I\wedge *\omega_J =
-{1\over 2}\partial_I\partial_J\log F(X)|_{F=1} .}

A solution \eqmsol\ of the equations of motion take the form:
\eqn\rtgflux{
G = {1\over \cal{V}}{1\over r^3}dt\wedge dr  \wedge \alpha^I \omega_I.
}
Here $r$ is a $D=5$ transverse radial coordinate and ${\cal{V}}$ is a
Clabi-Yau volume, and ``electric'' charges $\alpha_I = G_{IJ}\alpha^J $ are
$$
\alpha_{I}=\int_{C^{(4) I}\times S_3}(^{\ast} G)
$$
where $C^{(4) I}, \quad I = 1,..., h^{1,1}$ are 4-cycles in the
Calabi-Yau
manifold.
The nontrivial flux of this form leads to the appearance of the nonzero
scalar
potential. From the $G\wedge ^{\ast} G$ term of the  action it follows
that
\eqn\bhpottd{\eqalign{
\int_{M_{11}}G\wedge ^{\ast} G = \int_{M_{11}} {\sqrt{g_{str}}\over r^6}
{1
\over {\cal{V}}^2}\alpha^I\alpha^J\omega_I\wedge *\omega _J  =
2\int_{M_{5}}{\sqrt{g_{E}}\over r^6}{1\over
{\cal{V}}^2}\alpha_I\alpha_JG^{IJ}(X)
}}
here we use the relation $\sqrt{g_{str}}={\cal{V}}^{-{5\over
3}}\sqrt{g_{E}}$.
The effective potential for $D=5$ is:
$$
V_5={1\over r^6}{1\over {\cal V}^2}\alpha_{I}\alpha_{J}G^{IJ}(X).
$$

This D=5 potential $V_5$  explicitly depends  on  $r$  and fells with
the
distance to the source so that in this case there is no run-away
Calabi-Yau
volume as in  \refs{\luov, \begu}.

In the effective one-dimensional theory \dfactbh\ $U$ plays a role of a
dilaton
and the appearence of the effective potential \bhpotfo\ can be
understood as a
result of gauging of an additional axionic shift symmetry.

The consideration of Section 3 shows that hypermultiplets can also play
a very
important role in the physics of black holes. The difference between a
black
hole solution and a domain wall solution is in the presence of
non-trivial
vector fields in the black hole case. Those fields should be added to
the
supersymmetry variations \killingeqs\ (see \refs{\ceda}):
\eqn\killingeqsbh{\eqalign{
\delta\psi_m^{A}  = & D_m\epsilon^{A} + {i\over 8}X_I(\Gamma_m^{np} -
4\delta_m^n\Gamma^p)F^I_{np}\epsilon^{A} - {i \over 3}\,W_B^{\ A}\,
 \Gamma_m\epsilon^B \ , \cr
\delta \lambda^{Ai} =& -{i \over 2} \Big[ \Gamma^m \partial_m \phi^i\,
 \epsilon^{A} - 2 i\,   g^{ij} \partial_j W_B^{\ A}\, \epsilon^B \,
 \Big] +{3\over 8}g^{ij}\partial_j X_I \Gamma^{mn}F^I_{mn}\epsilon^{A} \
, \cr
\delta\zeta^\alpha =&  -{i \over \sqrt{2}}\, V^{A\alpha}_u\,\Big[
 \Gamma^m \partial_m q^u\, - 2  h^{uv} (J^x)_v^{\ r}
 \nabla_r W^x) \Big] \epsilon_A .
}
}

It is easy to see that in ungauged supergravity hypermultiplets are decoupled
from the theory and do not affect the solution. It may not be the case
when some gauging is present.

An application of Morse theory to the black hole physics can have interesting
consequences. It is possible to consider a black hole solution in gauged
supergravity with vector multiplet gauging \fdbhg\ . In this case
hypermultiplets are also decoupled and, unfortunately (as we will see
later, rather, predictably), the BPS solutions contain naked singularities or
blow up near the horizon. This means that in this theory it is impossible to
find a black hole with a regular horizon embedded in the $AdS_5.$ On the other
hand, application of Morse theory to this case predicts existence of only one
critical point (or several of the same type) and the absence of a second
non-trivial vacuum (see Section 5) is justified.

The case of the black hole solution with hypermultiplet gauging is much
more complicated. In this case, hypermultiplets are no longer decoupled from
the theory and the hypermultiplet gauging can lead to the appearance of
a second non-trivial critical point. It will be interesting to consider
such solutions in the future work.

%%%%%%%%%%%%%%%%%%%%%%%%%%%%%%%%%%%%%%%%%%%%%%%%%%%%%%%%

\newsec{Morse Theory and Vacuum Degeneracy}

%%%%%%%%%%%%%%%%%%%%%%%%%%%%%%%%%%%%%%%%%%%%%%%%%%%%%%%%%

In the previous sections we studied the effective dynamics of BPS
solutions in $\CN=2$ five-dimensional supergravity which preserve the
$SO(3,1)$ (or $SO(4)$) subgroup of the Lorentz symmetry, as well as
the effective dynamics of the similar solutions in four dimensions.
The examples --- such as supersymmetric domain walls or spherically
symmetric static black holes --- correspond to solutions where
space-time metric is a function of a single coordinate.  Effective
dynamics of a BPS state with these properties turns out to be
supersymmetric quantum mechanics of the form \morseaction, {\it cf.}
\refs{\brmi, \fegi, \den}.  The nature of the BPS state is perfectly
indifferent. For all such BPS states it is true that the solution
represents a gradient flow of the height function $h$ between two
critical points in the scalar field manifold $\CM$.  For example, in
five-dimensional gauged supergravity coupled to a certain number of
vector multiplets $\CM = \CM_V$ is a Riemannian manifold parameterized
by scalar fields $\phi^A$ from the vector multiplets and $h$ is
(logarithm of) the superpotential $W$.  Moreover, if this theory is
obtained from compactification of M theory on a Calabi-Yau three-fold,
then $\CM_V$ is just the K\"ahler structure moduli space of this
Calabi-Yau manifold, and $h$ has a microscopic interpretation in terms
of a $G$-flux \refs{\luov, \begu}.

In any case, the study of BPS objects described above and the
classification of supersymmetric vacua in $D=4$ and $D=5$ supergravity
boils down to a simpler problem in supersymmetric quantum mechanics
with the effective action \morseaction. This connection to supersymmetric
quantum mechanics allows one to address many interesting physical
questions studying more elementary system. For example, Klebanov
and Tseytlin used this relation to study supergravity duals of
the RG-flows in $SU(N) \times SU(N+M)$ gauge theories \KT.
In this section we will discuss another application of this relation.

%This will be our starting point in this section.
For a compact space $\CM$, it has been shown by Witten \witten\ that
supersymmetric quantum mechanics with the action \morseaction\ is just
Hodge-de Rham theory of $\CM$. Therefore, assuming it is also
the case for certain (non-compact) scalar field manifolds $\CM$
that occur in supergravity, we can use topological methods
--- in particular, Morse theory --- to classify possible BPS states
and supersymmetric vacua they connect.  As we explained above, our
results are quite generic irregardless of the physical nature of a
given BPS solution; they work equally well for BPS domain walls, for
spherically symmetric black holes, or any other BPS object with the
effective action \morseaction. We only make a couple of assumptions
necessary in the following.
First, we require $\CM$ to be a Riemannian manifold, although
sometimes it comes with some additional structure ({\it e.g.}
complex or quaternionic structure) corresponding to additional
(super-)symmetry in the problem.  One has to be very careful with this
assumption in the cases when $\CM$ has singularities, in particular,
in Calabi-Yau compactifications of M theory. Second, we assume that
the space $\CM$ is either compact or it has the right behaviour at infinity,
so that the topological methods of Morse theory are reliable\foot{For
an analog of Morse theory in complex non-compact geometry
that admits a holomorphic torus action see \NCMorse.}.

Given a Riemannian manifold $\CM$, let $h \colon \CM \to \R$
be a "good" Morse function of $C^{\infty}$-class.
This condition means that every point in $\CM$ is either
a regular point where $dh \ne 0$ or an isolated critical
point $p \in \CM$ where $dh =0$ and $h$ can be written as:
\eqn\localh{h (\phi^A) = h(p)
- \sum_{A=1}^k (\phi^A)^2 + \sum_{A=k+1}^n (\phi^A)^2}
in some neighborhood of $p$.

The number $k$ in \localh\ is called the Morse index of a critical
point $p$ and is denoted by $\mu (p)$.  It turns out that $\mu (p)$
has a nice physical interpretation in supergravity. Recall,
that the critical points of $h$ correspond to supersymmetric
vacua. Moreover, according to the analysis of section 2, the
eigenvalues of the Hessian of $h$ determine the type of the attractor
point.  In particular, critical points with zero Morse index are UV
attractive.  On the other hand, critical points with $\mu =n$ are IR
attractive.  In general, a critical point $p$ is of mixed type, $0 <
\mu <n$. In models with only vector multiplets, the superpotential
depends
generically on all scalars and such mixed critical points do not
correspond to stable vacua, see discussion after \scalar.

Therefore, in order to classify possible supersymmetric vacua,
one has to know how many points have a given Morse index.
Morse theory provides a nice answer to this problem in terms
of the topology of the scalar field manifold $\CM$.
Before we state the result let us introduce a few more notions
which will be convenient in the following.

As we explained in the previous sections, a supersymmetric domain
wall or a spherically symmetric black hole corresponds to
a gradient flow\foot{It is this place where we use the Riemannian
metric on $\CM$ to define a gradient vector field $\nabla h$.}
of $h$ from one critical point $p$ to another critical point $q$.
Consider the ``moduli space of these BPS solutions'' $M(p,q)$.
Namely, for a pair of critical points $p$ and $q$ we define:
$$
M(p,q) = \{ \phi \colon \R \to \CM ~\vert~
{d \phi \over dt} = - \nabla h,
\lim_{t \to - \infty} \phi (t) = p,
\lim_{t \to + \infty} \phi (t) = q \} / \sim
$$
to be the moduli space of the gradient trajectories from $p$ to $q$
modulo the equivalence relation $\phi(t) \sim \phi (t + {\rm const})$.
In general, $M(p,q)$ is not a manifold, though perturbing it a bit
we can always assume that it is a manifold (or a collection of points,
if we have a finite number of distinct gradient trajectories).

A classical result in Morse theory asserts that the real dimension
of $M(p,q)$ is given by:
\eqn\dimm{{\rm dim} M(p,q) = \mu(p) - \mu (q) -1.}
It immediately follows that non-singular BPS domain walls
(or spherically  symmetric black holes) of types (ii) and (iv)
interpolating between two vacua of the same kind do not exist
in theories where $h$ is a global Morse function,
{\it cf.} the discussion in the end of section 2.
Indeed, if there are two critical points of the same type
({\it i.e.} either both IR or both UV), then the virtual
dimension \dimm\ becomes negative.
Another physical result that follows from \dimm\ is that
domain walls of type (iii) connecting an IR attractive
point and a UV attractive point come in $(n-1)$-dimensional families.

It is curious to note that BPS solutions interpolating between
mixed (IR/UV) vacua also appear in Morse theory as boundary
components of $M(p,q)$. For example, codimension 1 boundary of
$M(p,q)$ consists of the points corresponding to two consequent
gradient flows: first from $p$ to some other critical point $s$,
$\mu (p) > \mu (s) > \mu (q)$, and then from $s$ to $q$:
$$
\partial M(p,q) = \cup_{\{ s \}}~ M(p,s) \times M(s,q).
$$
Similarly, codimension 2 boundary consists of the points
corresponding to three consequent flows $p \to s \to r \to q$
with $\mu (p) > \mu (s) > \mu (r) > \mu (q)$, {\it etc.}
Including all these boundary components
we obtain a compact oriented manifold $\overline M(p,q)$.

Now we define Witten complex $C_* (\CM, h)$ as a free abelian
group generated by the set of critical points of $h$:
$$
C_k (\CM, h) = \oplus_{\mu (p) = k} \IZ \cdot [p].
$$
Furthermore, we define
$\partial \colon C_k (\CM, h) \to C_{k-1} (\CM, h)$
via the sum over gradient trajectories (counted with signs):
$$
\partial [p] = \sum_{\mu (q) = k-1} \# M(p,q) ~ [q].
$$
Then, the Morse-Thom-Smale-Witten theorem says that
$\partial^2 = 0$ and:
\eqn\morseh{H_* (C_* (\CM, h)) = H_* (\CM, \IZ).}
Therefore, the classification of critical points and gradient
trajectories which represent supersymmetric vacua and
BPS solutions, respectively, can be addressed in terms
of the topology of $\CM$. In particular, one finds
the following lower bound on the total number of
supersymmetric vacua:
\eqn\morsech{\sum {\rm rank}~ C_i (\CM, h) \ge
\sum {\rm rank}~ H_i (\CM, \R).}
This is the classical Morse inequality.

We conclude that by using Morse theory formulas
\dimm\ -- \morsech\ one can classify
five-dimensional systems discussed in sections 2 and 3.
In what follows we illustrate these methods
in a number of interesting examples and, in particular,
we count the number of supersymmetric vacua computing
the Betti numbers $b_i = {\rm rank}~ H_i (\CM, \R)$
of the corresponding scalar field manifolds.
Even though the general formulas \dimm\ -- \morsech\ were derived for
a compact manifold $\CM$, the Morse inequality \morsech\ is
expected to hold in a wider class of examples, including certain
non-compact manifolds $\CM_V$ and $\CM_H$ relevant for supergravity.
%%%%%%%%%%%%%%%%%%%%%%%%%%%%%%%%%%%%%
%\bigskip
%\bigskip\noindent {\it Application to Vector Multiplets: a no-go
%theorem}
For example, let us consider $\CN \ge 2$ five-dimensional gauged
supergravity interacting with a certain number of vector multiplets.
As we will see in a moment, such theories possess at most one
supersymmetric critical point on every branch of the scalar field
manifold $\CM_V$, as long as interaction with other matter fields
({\it e.g.} hyper multiplets) can be consistently ignored,
and as long as the superpotential $W$ is generic enough
to be considered as a good global Morse function.

First, consider examples with exactly $\CN=2$ supersymmetry where
scalar fields $\phi^A$ take values in a real homogeneous cubic
hypersurface $\CM_V = \{ F=1 \}$ defined by \Fconstr\ in a vector
space parameterized by the fields $X^I$ with Minkowski signature \gusi.
We expect that real cubics of this form have trivial topology.
Even though we do not have  a mathematical proof of this fact,
we argue as follows.
Suppose, on the contrary, that $\CM_V$ is topologically non-trivial.
Then, there should exist at least two critical points,
one of which must have non-zero Morse index, $0 < \mu < n_V$.
However, as we explained in section 2, the existence of such points
would violate the c-theorem \frgu\ and contradict our original assumtion
that $\CM_V$ is a Riemannian manifold with a positive-definite metric.

Now let's see that scalar field manifold $\CM_V$ in $\CN>2$
five-dimensional gauged supergravity is also topologically trivial.
This result immediately follows from the fact\foot{We thank
R.~Kallosh and N.~Warner for explaining this to us.} that in
supergravity theories with more supersymmetry the space $\CM_V$ can
always be represented as a quotient space of a non-compact
group $G$ by its maximal compact subgroup $H$ \FGK.
Note, that $G$ can have time-like directions which would
be inconsistent if we did not divide by $H$ that makes
the metric on the quotient space $\CM_V = G/H$ positive-definite.
Another effect of the quotient by $H$ is that the space
$G/H$ is topologically trivial unless we divide further
by a discrete symmetry group, {\it e.g.} U-duality
group\foot{We are grateful to E.~Witten for pointing out that a quotient
by a discrete group may lead to interesting non-trivial topology of $\CM_V$.}.
This is one way to make the topology of $\CM$ topologically non-trivial.

%There are a few remarks in place here.
%First of all, the no-go theorem applies to theories with
%vector multiplets only. Once we allow interaction with
%other types of matter fields, there may be critical points
%with arbitrary Morse index and, therefore, the total space
%of scalar fields, $\CM$, may be topologically non-trivial.
%For a simple example see \frgu\ where a theory of one
%vector and one hyper multiplet with the target space
%$\CM = {SU(2,1) \over SU(2) \times U(1)} \times SO(1,1)$
%was shown to have a supersymmetric saddle point.
%A negative eigenvalue of the Hessian of $W$ at this point
%corresponds to turning on the scalar fields in the hyper multiplet.

The second possibility to get theories with multiple supersymmetric
vacua is to take a space of scalar fields $\CM$ with more than one branch.
For instance, an important family of theories is based on
the target spaces of the following simple form:

$\underline{\CM_V=SO(n,1)/SO(n)}:$ Since $SO(n)$ is the maximal
compact subgroup in the non-compact group $SO(n,1)$ the quotient
space $\CM_V$ is equivalent, up to homotopy, to a set two points.
The number of points comes from the number of disconnected
components in the non-compact group $SO(n,1)$ which looks like
a hyperbolic space. Therefore, irregardless of the value of $n$
in this class of models we find the following Betti numbers
$b_i = {\rm rank}~ H_i (\CM_V, \R)$:
$$
b_0 = 2,
$$
$$
b_i =0, \quad i>0.
$$
{}From the Morse inequality \morsech\ it follows that
the corresponding supergravity theories with generic $W$
possess at least two UV attractive supersymmetric vacua.
By numerical computations, one can verify that there are
exactly two vacua of UV type. We remark that there are
no smooth domain walls because the two vacua belong
to different disconnected branches.

$\underline{\CM_V=SO(n-1,1) \times SO(1,1) /SO(n-1)}:$
Once again, in this example we divide a non-compact group
by its maximal compact subgroup, so that the resulting space
is isomorphic to a set of points for the reason explained above.
This time we get 4 points, one for every disconnected component
of $SO(n-1,1) \times SO(1,1)$. For the Betti numbers of $\CM_V$
we obtain:
$$
b_0 = 4,
$$
$$
b_i =0, \quad i>0.
$$
Similar to the previous example, there are at least four
UV attractive vacua which can not be smoothly connected by
BPS domain walls.

These are examples of symmetric spaces which typically
appear as scalar field manifolds in  $\CN=2$ five-dimensional
supergravity theories and also in models with additional
(super-)symmetry structure \groupsref, as we mentioned earlier.

It is important to stress here that it is crucial for the height
function $h$ to be globally defined over the entire target space $\CM$.
For example, this assumption breaks down in a very important class
of models corresponding to M theory compactification on Calabi-Yau
three-folds. In these models $\CM_V$ is just the K\"ahler structure
moduli space of the Calabi-Yau manifold.  Since the K\"ahler structure
moduli spaces usually have trivial topology, one might naively
conclude from \morseh\ that there is only one vacuum (a UV attractive
fixed point) and no non-trivial domain walls. However, in general,
$\CM_V$ consists of several K\"ahler cones separated by the walls where
certain algebraic curves in the Calabi-Yau space shrink to zero size.
Local anomaly arguments in heterotic M theory suggest that $G$-flux
should jump while crossing a K\"ahler wall \greene.  In fact, passing
through a flop transition point the second Chern class changes and,
therefore,
the 5-brane charge induced in the boundary field theory also jumps.
Since the total 5-brane charge should be conserved (and equal to zero
in the compact Horava-Witten setup) some $\alpha_I$ also have to jump,
as if the flop curves effectively carry a magnetic charge
\refs{\greene}.  Although the superpotential \superpotn\ passes these
curves smoothly, because the corresponding $X^I$ vanishes there,
its (second) derivatives jump once we cross a K\"ahler wall,
and the corresponding $h$ is {\it not} a globally defined height
function.

%%%%%%%%%%%%%%%%%%%%%%%%%%%%%%%%%%%%%%%%%%%%%
%\bigskip\noindent {\it Application to Hyper Multiplets}

A similar result occurs in  gauged supergravity theories coupled to hyper multiplets.
In this case supersymmetry implies that scalar fields from hyper multiplets
parameterize a quaternionic manifold $\CM_H$ of negative curvature \BW:
$$
R = -8 (n_H^2 + 2n_H)
$$
Even though this condition is not as much restrictive as the supersymmetry
condition \secondder\ in the case of vector multiplets, known examples of
quaternionic homogeneous coset spaces that may serve as hypermultiplet
target manifolds are typically non-compact and topologically trivial:

%Now let us discuss some examples of quaternionic homogeneous
%coset spaces that may serve as a hypermultiplet target space, $\CM_H$.

$\underline{\CM_H=SO(n,4)/SO(n) \times SO(4)}:$ Similar
to the example $\CM=SO(n,1)/SO(n)$, this quaternionic
space is contractible since\foot{Stricty speaking,
$S(O(n) \times O(4))$ is the maximal compact
subgroup of $SO(n,4)$.} $SO(n) \times SO(4)$ is
the maximal compact subgroup of $SO(n,4)$. Therefore,
$\CM_H$ is homologous to a set of two points:
$$
b_0 = 2,
$$
$$
b_i =0, \quad i>0.
$$

The same result we find for $\CM_H=SU(n,2)/SU(n) \times SU(2)$.
As we mentioned earlier, a simple way to obtain models where scalar
fields take values in a topologically non-trivial manifold $\CM_H$ is to
devide by a discrete group which, for example, may be a subgroup
of the isometries of $\CM_H$. For example, if we have only the universal
hypermultiplet, $n=1$, the coset space $\CM_H=SU(1,2)/U(1) \times SU(2)$
is a quaternionic K\"ahler manifold, where the K\"ahler potential
can be written as $K (S,C) = - \log (S + \bar S - 2(C + \bar C)^2)$.
It has two abelian isometries corresponding to shifts of the complex
scalar fields $S$ and $C$, $S \to S + i a$ and $C \to C + i b$.
To get a topologically non-trivial manifold, we can consider a quotient space:
$$
\CM_H / \Z^2 =  { SU(1,2)  \over U(1) \times SU(2) \times \Z^2}
$$
where the action of $\Z^2$ is equivalent
to identification $S \sim S + i a$ and $C \sim C + i b$ for integer
numbers $a$ and $b$. Supergravity theory coupled to a hypermultiplet
based on the resulting quotient space is expected to have at least
$\sum_i b_i (\CM_H / \Z^2) = 4$ supersymmetric vacua.

Before we conclude this section, let us remark that due to its construction,
the superpotential or height function may not depend on all hyper-scalars
and the chosen gauged isometry of ${\CM_H}$ determines the scalars on
which the superpotential depends.  Hence, for a given superpotential,
obtained by a specific gauging, a mixed critical point with $0<\mu < n_H$
may appear as a ``good'' UV or IR fixed point. Only the critical
points with $\mu=0$ and $\mu=n_H$ are ``gauge-independent''
and appear in all gaugings as UV and IR fixed points.
Assuming that Morse inequalities are saturated, every component of
$\CM$ has exactly one UV critical point with $\mu =0$ and at
most one IR point with $\mu =n_H$ if $\CM$ is a compact manifold\foot{It
should be stressed, however, that the authors do not know whether
supergravity theories based on compact scalar field manifolds $\CM$
exist or not.}.
The values of all scalar fields are fixed in these two critical points.
Additional (mixed) critical points can be stable under the UV/IR
scaling only if the superpotential has (bad) flat directions.

%%%%%%%%%%%%%%%%%%%%%%%%%%%%%%%%%%%%%%%%%%%%%%%%%%%

\newsec{Examples: $SL(3)$ Symmetric Coset Spaces}

%%%%%%%%%%%%%%%%%%%%%%%%%%%%%%%%%%%%%%%%%%%%%%%%%%%%%

Let us consider in more detail a specific example
where the quotient space which appears in $\CN=2$
five-dimensional supergravity is associated with Jordan
algebras of the form:
$$
\CM = { \rm{Str}_0 (J) \over {\rm Aut} (J)}
$$
where $\rm{Str}_0 (J)$ is the reduced structure group and ${\rm Aut}
(J)$ is the automorphism group of a real unital Jordan algebra of
degree 3.  A simple example based on irreducible $J$ is
$\CM=SL(3,\R)/SO(3)$, which is a three-dimensional analog of the
Lobachewsky plane, $SL(2,\R)/SO(2)$.  Like its two-dimensional analog,
the space $\CM$ is contractible because a semisimple Lie group
$SL(n,\R)$ is isomorphic to its maximal compact subgroup
$SO(n)$. After we divide by the latter we get a point, up to homotopy:
\eqn\pointtop{b_0 = 1}
$$
b_i =0, \quad i>0.
$$
So, we come to the conclusion that the STU model with generic
superpotential has a single supersymmetric vacuum.
In the rest of this section our goal will be to identify
the physics of this vacuum.

There are two generalizations of this example over complex numbers and
quaternions. In any case we divide $SL(3,F)$,
where $F = \{ \R, \C, \H \}$ is the base field in question, by its
maximal compact subgroup.  Although the resulting quotient space has
trivial topology \pointtop\ for all ground fields $F$, the
physics is different.  Namely, we claim that three supersymmetric
vacua corresponding to various choices of $F$ describe AdS$_5$
dual of $\CN=4$ super-Yang-Mills perturbed by mass terms which
preserve different subgroups of $SO(6)$ R-symmetry.

In order to see that a scalar vev. gives rise to a mass deformation,
one has to find scaling dimension $\Delta^{(i)}$ of the corresponding
operator $\CO^{(i)}$ in the boundary theory.
Using the standard formula \massformel\ we get $\Delta^{(i)} =2$
which allows us to identify $\CO^{(i)}$ with a mass term.
In general, a mass term is specified by a bosonic symmetric
$6\times 6$ matrix and a fermionic symmetric $4\times 4$ matrix.
In our three exampes, however, these matrices have additional
symmetries reminiscent of $\CN=4$ super-Yang-Mills theory
with R-symmetry twists \ganor.
In particular, at special values of twists,
where the corresponding phases are equal to $(-1)$,
extra hypermultiplets become massless\foot{To see this,
it is convenient to think of gauge theories with R-symmetry
twists as compactifications of five-dimensional gauge theories
on a circle with twisted boundary conditions on $S^1$ \ganor.
The masses of Kaluza-Klein states are given by $m= n + 1/2 +  q W$
where $W$ is the value of the Wilson lines of
the gauge field and $q$ is the charge of a given mode.
Note that for any $n \in \Z$ we get a whole hyper-multiplet.
Moreover, for $n=0$ and $-1$ and $W=1/2$ and $q=+1$ and $-1$
we find two massless hypermultiplets.},
and we expect to get a four-dimensional superconformal theory
\foot{We thank Ori Ganor for discussions on this point.}.

Below we discuss in more details the case where $F = \C$.
In particular, we solve the flow equations \bps\ for the coset
space ${SL(3,\C) / SU(3)}$.
This coset manifold is parameterized by 8 non-trivial scalar
fields with the intersection form \refs{\gusii}
\eqn\prepot{
F = STU - S |X|^2 - T |Y|^2 - U |Z|^2 + 2 \, {\rm Re}(XYZ)
}
where $X,Y$ and $Z$ are complex and $S$, $T$ and $U$ are real. The
unique solution of the attractor equations \refs{\fekaa}: $\partial_I F
= e^{-2U} \, H_I$, which solves the flow equations \bps\ (see
\refs{\begu}) is given
\eqn\solution{
\eqalign{
S = (H_T H_U - { 1 \over 4} |H_X|^2 ) \; e^{-4U} \ ,
\qquad & X = {1 \over 4} (\bar H_Y \bar H_Z - 2 H_X H_S ) \; e^{-4U} \ ,
\cr
T = (H_S H_U - { 1 \over 4} |H_Y|^2 ) \; e^{-4U} \ ,
\qquad & Y = {1 \over 4} (\bar H_X \bar H_Z - 2 H_Y H_T ) \; e^{-4U} \ ,
\cr
U = (H_S H_T - { 1 \over 4} |H_Z|^2 ) \; e^{-4U} \ ,
\qquad & Z = {1 \over 4} (\bar H_X \bar H_Y - 2 H_Z H_U ) \; e^{-4U}
}
}
where $H_I$ is a set of harmonic functions
$$
H_I = h_I + 6\, \alpha_I \, y
$$
which are real for the $S,T,U$ components and complex for the $X,Y,Z$
components. For the metric we take the ansatz:
\eqn\metric{
ds^2 = e^{2U} \Big[ -dt^2 + d\vec{x}^2 \Big] + e^{-4U} dy^2 \
}
where the function $e^{-2U}$ is obtained from the requirement $F=1$:
\eqn\ufunct{
e^{6U} = H_S H_T H_U - {1 \over 4} \Big( H_S |H_X|^2 + H_T |H_Y|^2 +
H_U |H_Z|^2 \Big) + {1 \over 4} \, {\rm Re}(H_X H_Y H_Z) \ .
}
The scalar fields are defined by $F=1$ and we may consider,
for example, the ratios: $\phi^A = \{{T \over S} , {U \over S} , {X
\over S} , {Y\over S} , {Z \over S}\}$.  The uniqueness of the
solution fits very well with our expectation from Morse theory,
that this coset allows only a unique critical point.
At the critical point the space time becomes $AdS_5$ as
$y \to + \infty$ with the negative cosmological constant:
$$
\Lambda  = - (e^{4U}/y^2)_{y \rightarrow +\infty}
$$

In the case of a unique critical point with a negative cosmological
constant, it is natural to ask what is the four-dimensional
field theory dual to this AdS$_5$ vacuum.

Notice that the scalar fields $\phi^A$
defined as ratios of harmonic functions stay finite in the AdS vacuum
and approach their critical values.  Having the explicit solution, we
can also calculate the supergravity effective action. As given by
\bogom, the BPS nature ensures vanishing of the bulk term and the
surface part yields\foot{Note the different coordinate system.}:
\eqn\action{
S_{eff} = {2 \over 3} \Big(\partial_y e^{6U}\Big)_{y = + \infty} = \Big(
-2
\Lambda y^2 + a_1 \, y\Big)_{y \rightarrow \infty} + a_2. }
As expected, this action has singular terms and a finite part.  The
leading singularity scales with the cosmological constant (AdS volume)
while the subleading term $a_1$ and the finite term $a_2$ can be
obtained by inserting the harmonic functions \ufunct.  The divergent
part will be subtracted by the renormalization in the field theory and
the finite part will give the renormalized effective action;
in our approximation we see only potential or mass terms.

{From} the RG-flow point of view this AdS vacuum corresponds to an UV
fixed point. While moving towards negative $y$ the warp factor
$e^{2U}$ decreases monotonically (according to $c$-theorem)
and we approach the IR region in the fields theory.
Because $e^{6U}$ is negative at $y= - \infty$,
we have to pass a zero at some finite value of $y$,
which is the singular end-point of the RG flow.
Like in any other case with vector multiplets only,
the absence of an IR fixed point forces the solution to
run into a singularity.

On the other hand, from the string theory perspective this solution
corresponds to a 5-brane wrapping a holomorphic 2-cycle,
namely a torus $T^2$.
Once again, we point out the analogy with the construction
of gauge theories with R-symmetry twists from (2,0) theory
on a torus in the limit where the size of the torus
and the values of twists tend to zero (${\rm Vol} (T^2) \to 0$
and $\alpha \to 0$) while their ratio $\alpha / {\rm Vol} (T^2)$
remains finite and defines a mass scale in the resulting theory \ganor.
{From} this construction
it is clear that at the special values of twists $\alpha$,
where the theory becomes superconformal, it must be
dual to AdS$_5 \times S_5$ perturbed by a dimension-8
operator (proportional to ${\rm Vol} (T^2)$) and dimension-2
operators (proportional to $\alpha$), similar to
the AdS$_5$ vacuum of $SL(3)$ coset spaces we found.

It is reasonable to put another (5-brane) source at some
place where the warp factor is still positive, say $y=0$.
This extra source appears as a
non-trivial right-hand side of the harmonic equations:
$$
\partial^2 H_I \sim \alpha_I \delta(y)
$$
where $\alpha_I$ is the component of the 5-brane charge related to a
basis of $(1,1)$-forms $\omega_I$.  There are two possibilities:
in the first option we continue in a symmetric way through the source,
which implies the replacement
$H_I \rightarrow h_I + 6 \alpha_I \, |y|$ and is
equivalent to a sign change in the flux vector $\alpha_I$ while
passing the source at $y=0$.  This case appears in the $\Z_2$ orbifold
of the Horava-Witten setup compactified to 5 dimensions
\refs{\luov}. But we may also consider the case where the flux jumps
from zero to a finite value, i.e.\ on the side behind the source we can
set $\alpha_I=0$, which is equivalent to the replacement $H
\rightarrow h_I + 6 \alpha_I \, {1 \over2} (y + |y|)$,
so that $H_I$ is constant for negative $y$ and the space time is flat.

By adding this source, we cut off the (singular) part of space-time
and glue instead an identical piece ($Z_2$ symmetric) or flat space
(vanishing flux on one side).  If one wants to discuss a RS-type
scenario, one may also cut off the regular AdS part and keep on both
sides the naked singularity.  In the first case the source is the
standard positive tension brane generating an AdS space on both sides,
whereas in the second case a negative tension brane is accompanied
with a naked singularity. For a more detailed discussion of sources see
\kbtsing .

\vskip 30pt

%%%%%%%%%%%%%%%%%%%%%%%%%%%%%%%%%%%%%%%%%%%%%%%%%%%%%%

\centerline{\bf Acknowledgments}

We are grateful Alexander Chervov, Ori Ganor, Brian Greene, Renata Kallosh,
Eric Sharpe, Gary Shiu, Andrew Strominger, Nicholas Warner,
and Edward Witten for helpful discussions and comments.
The work of K.B.\ was partly done at the Theory group of Caltech and is
supported by a Heisenberg grant of the DFG and by the European
Commission RTN programme HPRN-CT-2000-00131.
S.G. is supported in part by the Caltech Discovery Fund,
grant RFBR No 98-01-00327 and Russian President's grant No 00-15-99296.

\listrefs
\end